\documentclass[iop,numberedappendix]{emulateapj}

\usepackage{txfonts}
\usepackage{graphicx}
\usepackage{subfigure}
\usepackage{multirow}
\usepackage{verbatim}
\usepackage{longtable}
\usepackage{url}
\usepackage{float}
\usepackage{ esint }
\usepackage{dcolumn}
\usepackage{hyperref}
\usepackage[utf8]{inputenc}
\makeatletter
\newcommand\footnoteref[1]{\protected@xdef\@thefnmark{\ref{#1}}\@footnotemark}
\makeatother

\slugcomment{Accepted to AJ}

\shorttitle{Robo-AO \textit{Kepler} Planetary Candidate Survey IV}
\shortauthors{Ziegler et al.}

\begin{document}

\title{Robo-AO Kepler Survey IV: the effect of nearby stars on 3857 planetary candidate systems}

\author{Carl Ziegler\altaffilmark{1}, Nicholas M. Law\altaffilmark{1}, Christoph Baranec\altaffilmark{2}, Reed Riddle\altaffilmark{3}, Dmitry A. Duev\altaffilmark{3}, Ward Howard\altaffilmark{1}, Rebecca Jensen-Clem\altaffilmark{4}, S. R. Kulkarni\altaffilmark{3}, Tim Morton\altaffilmark{5}, Ma{\"i}ssa Salama\altaffilmark{2}}

\email{carlziegler@unc.edu}
\altaffiltext{1}{Department of Physics and Astronomy, University of North Carolina at Chapel Hill, Chapel Hill, NC 27599-3255, USA}
\altaffiltext{2}{Institute for Astronomy, University of Hawai`i at M\={a}noa, Hilo, HI 96720-2700, USA}
\altaffiltext{3}{Department of Astronomy, California Institute of Technology, 1200 E. California Blvd., Pasadena, CA 91101, USA}
\altaffiltext{4}{University of California, Berkeley, 510 Campbell Hall, Astronomy Department, Berkeley, CA 94720, USA}
\altaffiltext{5}{Department of Astrophysical Sciences, Princeton University, Princeton, NJ 08544, USA}

\begin{abstract}
We present the overall statistical results from the Robo-AO \textit{Kepler} planetary candidate survey, comprising of 3857 high-angular resolution observations of planetary candidate systems with Robo-AO, an automated laser adaptive optics system. These observations reveal previously unknown nearby stars blended with the planetary candidate host star which alter the derived planetary radii or may be the source of an astrophysical false positive transit signal. In the first three papers in the survey, we detected 440 nearby stars around 3313 planetary candidate host stars. In this paper, we present observations of 532 planetary candidate host stars, detecting 94 companions around 88 stars; 84 of these companions have not previously been observed in high-resolution. We also report 50 more-widely-separated companions near 715 targets previously observed by Robo-AO. We derive corrected planetary radius estimates for the 814 planetary candidates in systems with a detected nearby star. If planetary candidates are equally likely to orbit the primary or secondary star, the radius estimates for planetary candidates in systems with likely bound nearby stars increase by a factor of 1.54, on average. We find that 35 previously-believed rocky planet candidates are likely not rocky due to the presence of nearby stars. From the combined data sets from the complete Robo-AO KOI survey, we find that 14.5$\pm0.5\%$ of planetary candidate hosts have a nearby star with 4\arcsec, while 1.2$\%$ have two nearby stars and 0.08$\%$ have three. We find that 16$\%$ of Earth-sized, 13\% of Neptune-sized, 14$\%$ of Saturn-sized, and 19$\%$ of Jupiter-sized planet candidates have detected nearby stars.
\end{abstract}


\section{Introduction}

Over its initial four year mission, the \textit{Kepler} telescope observed hundreds of thousands of stars, searching for the slight dip in brightness consistent with a transiting exoplanet. \textit{Kepler} has exquisite photometric precision but relatively low spatial resolution, with an effective point-spread function of 6-10\arcsec and a pixel size of $\sim$4\arcsec \citep{haas10}. The majority of \textit{Kepler} targets are solar-type \citep{batalha13}, and most form with at least one companion star \citep{duquennoy91, raghavan10}.  These companion\footnote{For brevity we denote stars which we found within our detection radius of KOIs as ``companions,'' in the sense that they are asterisms associated on the sky.} stars are often blended with the planetary host star in the \textit{Kepler} aperture, resulting in inaccurate host star characterization \citep{dressing13, santerne13} and a high number of astrophysical false positive transit signals, estimated to be $\sim$10\% of planetary candidates \citep{morton11, fressin13}.  Even when the candidates are bona fide planets, the planet radius measurements based on the diluted transit signal are underestimated due to the presence of multiple stars in the system or unbounded stars within the \textit{Kepler} photometric aperture \citep{morton11}.  All planetary candidates discovered with light curves produced by \textit{Kepler} must, therefore, be independently validated by ground-based high-angular resolution observations.

The challenge of performing high-angular resolution follow-up observations of the 4100 planet candidates (\textit{Kepler} objects of interest, or KOIs) discovered by \textit{Kepler} \citep{borucki10, borucki11a, borucki11b, batalha13, burke14, rowe14, coughlin15, morton16, dr25} has been met with considerable effort by the community \citep{howell11, adams12, adams13, lillo12, lillo14, horch12, horch14, marcy14, dressing14, gilliland15, wang15a, wang15b, torres15, everett15, kraus16, furlan16}. Many of these surveys were performed with large-aperture telescopes, sensitive to close (tens of mas separation) and faint (8-10 magnitudes fainter than the host star) nearby stars. However, the combined efforts of surveys with traditional high-resolution instruments--in particular, adaptive optics--has resulted in a piecemeal approach, covering less than half of the KOIs. This is in part a result of redundant observations of a small set of KOIs, as the target lists of these surveys are often biased towards bright stars. This bias also results in a high fraction of early-type stars and stars closer to the Sun, which skews any interpretations drawn from the data.  In addition, disparities in the instruments and passbands of these observations may lead to inconsistent vetting as each survey has different detection sensitivities to nearby stars. The comprehensive statistics and correlations that can be derived from a homogeneous dataset of thousands of high-resolution images of multiple stellar systems hosting planets are extremely difficult to  when using data from multiple surveys.

 With an order-of-magnitude increase in observational time-efficiency compared to traditional systems provided by Robo-AO, the first fully automated laser adaptive optics system, we are performing high-resolution imaging of every KOI system to search for companions with separations between 0\farcs15 and 4\farcs0. The first paper in this survey, \citet[hereafter Paper I]{law14}, observed 715 \textit{Kepler} planetary candidates, identifying 53 companions, with 43 new discoveries, for a detected companion fraction of 7.4\%$\pm$1.0\% within separations of 0$\farcs$15 to 2$\farcs$5.  The second paper in this survey, \citet[hereafter Paper II]{baranec16}, observed 969 \textit{Kepler} planetary candidates, identifying 202 companions, with 139 new discoveries, for a detected companion fraction of 11.0\%$\pm$1.1\% within separations of 0$\farcs$15 to 2$\farcs$5, and 18.1\%$\pm$1.3\% within separations of 0$\farcs$15 to 4$\farcs$0.  The third paper, \citet[hereafter Paper III]{ziegler16} in this survey observed 1629 KOIs, around which 223 companions were found around 206 KOIs, for a detected companion fraction of 12.6\%$\pm$0.9$\%$ within 4$\farcs$0 of planetary candidate hosting stars.

This paper presents the detection of nearby stars from observations of 532 KOIs, as well as expands the search for nearby stars around 715 KOIs observed initially in Paper I from its initial separation limit of 2\farcs5 to 4\farcs0. We also present the cumulative statistics from the survey, as well as derive corrected planetary radii for every candidate planet in a system with an observed nearby star.

We begin in Section \ref{sec:targetselection} by describing our target selection, the Robo-AO system, and follow-up observations. In Section \ref{sec:datareduction} we describe the Robo-AO data reduction and the companion detection and analysis. In Section \ref{sec:Discoveries} we describe the results of this survey, including discovered companions, and compare to other KOI surveys.  We discuss the results in Section \ref{sec:Discussion} and conclude in Section \ref{sec:conclusion}.

\section{Survey Targets and Observations}
\label{sec:targetselection}

\subsection{Target Selection}
\begin{figure*}
\centering
\includegraphics[width=0.83\paperwidth]{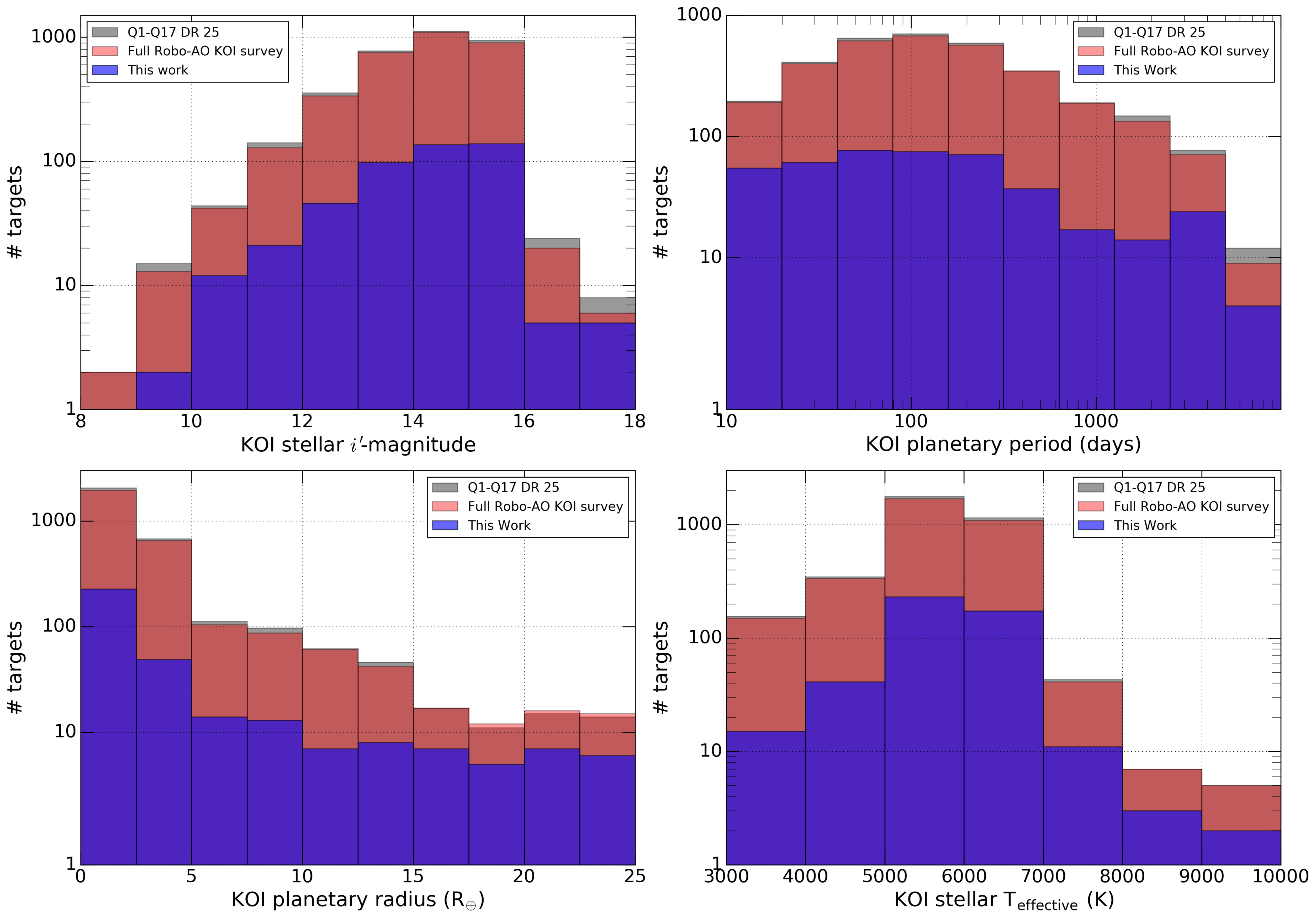}
\caption{Comparison of the distribution of new Robo-AO observations in this paper as well as the combined Robo-AO survey (Paper I, Paper II, Paper III, and this work) to the complete set of KOIs from Q1-Q17 \citep{borucki10, borucki11a, borucki11b, batalha13, burke14, rowe14, coughlin15, dr25}. Some observed KOIs with `CANDIDATE' disposition in early data releases that were observed with Robo-AO have been modified to `FALSE POSITIVE' in later releases, leading to a higher number of targets observed in some parameter bins, specifically at large planetary radii, than there are candidate systems in the latest \textit{Kepler} data release.}
\label{fig:histograms}
\end{figure*}

The objective of the Robo-AO \textit{Kepler} survey is to perform high-resolution observations of every KOI. We therefore targeted KOIs not observed in Paper I, Paper II, and Paper III from the \textit{Kepler} DR25 catalog based on Q1-Q17 data \citep{borucki10, borucki11a, borucki11b, batalha13, burke14, rowe14, coughlin15, dr25}. Observations of these targets presented in this paper are from the 2016 observing season.  KOIs flagged as false positives using \textit{Kepler} data were removed. In Figure$~\ref{fig:histograms}$, the properties of the targeted KOIs in this work as well as for the full Robo-AO survey as of the end of the 2016 observing season are compared to the set of all KOIs from Q1-Q17 with CANDIDATE dispositions based on \textit{Kepler} data. The Robo-AO \textit{Kepler} survey has observed more than 95\% of KOIs, and the distribution of observed KOIs in the survey closely matches the full KOI list in magnitude, planetary radius, planetary orbital period, and stellar temperature.

To compile a homogeneous survey, the observations of 715 KOIs in Paper I were re-analyzed to search for companions between the 2\farcs5 separation limit implemented in that paper and the 4\farcs0 separation limit of Papers II and III.  Observations of these targets were performed in the 2012 observing season.

\subsection{Observations}
We obtained high-angular-resolution images of 532 KOIs not previously observed by Robo-AO during 18 separate nights of observations between 2016 June 08 and 2016 July 15 (UT), detailed in Table$~\ref{tab:whitelist}$ in the Appendix.  The observations were performed using the Robo-AO laser adaptive optics system \citep{baranec13, baranec14, riddle12} mounted on the Kitt Peak 2.1-m telescope \citep{jensenclem17}, masked to a 1.85-m aperture. The AO system runs at a loop rate of 1.2 kHz to correct high-order wavefront aberrations.  Observations were taken in a long-pass filter cutting on at 600 nm (LP600 hereafter).  The LP600 filter approximates the \textit{Kepler} passband at redder wavelengths, while also suppressing blue wavelengths that reduce adaptive optics performance. The LP600 passband is compared to the \textit{Kepler} passband in Figure 1 of Paper I.

Typical seeing at the Kitt Peak Observatory is between 0$\farcs$8 and 1$\farcs$6, with a median around 1$\farcs$3 \citep{jensenclem17}. The typical FWHM (diffraction limited) resolution of the Robo-AO system is 0$\farcs$15. Images are recorded on an electron-multiplying CCD (EMCCD), allowing short frame rates for tip and tilt correction in software using a natural guide star ($m_V < 16$) in the field of view.  Specifications of the entire Robo-AO KOI survey are summarized in Table$~\ref{tab:specs}$.

In addition to new observations, we also search for companions at wider separations from the Paper I target list, using observations taken with Robo-AO at Palomar Observatory. The description of these observations is available in Section 2.2 of Paper I, with the full target list available in Table 5 of Paper I.

\begin{table}
\renewcommand{\arraystretch}{1.3}
\begin{longtable}{ll}
\caption{\label{tab:survey_specs}The specifications of the Robo-AO KOI survey}
\\
\hline
KOI targets    	& 3857 \\
FWHM resolution   	& $\sim$0$\farcs$15 (@600-750 nm) \\
Observation wavelengths & 600-950 nm\\
Detector format & 1024$^2$ pixels\\
Pixel scale & 43 mas/pix (Palomar)\\
& 35 mas/px (Kitt Peak)\\
Exposure time & 90 seconds \\
Targets observed / hour & 20\\
Observation dates & 2012 July 16 --\\at Palomar &  2015 June 12\\

Observation dates & 2016 June 8 --\\at Kitt Peak &  2016 July 15\\
\hline
\label{tab:specs}
\end{longtable}
\end{table}

\section{Data Reduction}
\label{sec:datareduction}
With a large adaptive optics dataset acquired by Robo-AO, the data reduction process was automated as much as possible for efficiency and consistency.  As in previous papers in the survey, after initial pipeline reductions described in Section \ref{sec:pipeline}, the target stars were identified (Section \ref{sec:targetverification}), PSF subtraction performed (Section \ref{sec:psfsubtraction}), nearby stars identified by visual inspection and by an automated companion search algorithm (Section \ref{sec:compsearch}), and constraints of the nearby star sensitivity of the survey measured (Section \ref{sec:imageperf}).  Finally, the properties of the detected companions are measured in Section \ref{sec:characterization}.

\subsection{Imaging Pipeline}
\label{sec:pipeline}

The Robo-AO imaging pipeline \citep{law09, law14} reduced the images: the raw EMCCD output frames are dark-subtracted and flat-fielded and then stacked and aligned using the Drizzle algorithm \citep{fruchter02}, which also up-samples the images by a factor of two.  To avoid tip/tilt anisoplanatism effects, the image motion was corrected by using the KOI itself as the guide star in each observation.

\subsection{Target Verification}
\label{sec:targetverification}

To verify that the star viewed in the image is the desired KOI target, we created Digital Sky Survey and UKIRT \citep{lawrence07} cutouts of similar angular size around the target coordinates. Each image was manually checked to assure no ambiguity in the target star and images with either poor performance or incorrect fields were removed.  These bad images made up approximately 1$\%$ of all our images, and for all of the targets additional images were available. 

 We select a 4\arcsec separation cutoff for our companion search to detect all nearby stars that would blend with the target KOI in a \textit{Kepler} pixel. To facilitate the automation of the data reduction, centered 8$\farcs$5 square cutouts were created around the 532 verified target KOIs, slightly larger than the diameter of our adopted separation limit so as to not remove a portion of the PSF of any nearby star within 4\arcsec.

\subsection{PSF Subtraction}
\label{sec:psfsubtraction}

To identify close companions, a custom locally optimized point spread function (PSF) subtraction routine based on the Locally Optimized Combination of Images algorithm \citep{lafreniere07} was applied to centered cutouts of every star. The code uses a set of twenty KOI observations, selected from the observations within the same filter closest to the target observation in time, as reference PSFs. We address the potential that nearby stars will not be detected due to the use of other KOIs as reference images in Section \ref{sec:psfcollisions} of the Appendix, and find that no nearby stars are likely to be missed. A locally optimized PSF is generated and subtracted from the original image, leaving residuals consistent with photon noise. This procedure was performed on all KOI images out to a radius of 2$\arcsec$ from the host star. Figure 4 in Paper III shows an example of the PSF subtraction performance.

\subsection{Companion Detection}
\label{sec:compsearch}

An initial visual companion search on the original and PSF-subtracted images was performed redundantly by two of the authors. This search yielded a preliminary companion list and filtered out bad images.

Continuing the companion search, we ran all images through a custom automated search algorithm, based on the code described in Paper I. The algorithm slides a 5-pixel diameter aperture within concentric annuli centered on the target star. For each annulus, the mean and standard deviation of the local noise is estimated using the fluxes within these apertures, with a sigma clip employed to remove any anomalously high signals such as those arising from a real astrophysical source. Any aperture with a summed signal greater than +5$\sigma$ compared to the local noise is considered a potential astrophysical source. These are subsequently checked manually, eliminating spurious detections with dissimilar PSFs to the target star and those having characteristics of a cosmic ray hit, such as a single bright pixel or bright streak.  The detection significance of detected companions are listed in Tables$~\ref{tab:paper1_table}$ and $~\ref{tab:newkois_table}$.

\subsection{Imaging Performance Metrics}
\label{sec:imageperf}

The two dominant factors that affect the image performance of the Robo-AO system are seeing and target brightness.  An automated routine was used to classify the image performance for each target. The code uses PSF core size as a proxy for image performance.  Observations were binned into three performance groups, with 31\% fall in the low-performance group, 41\% in the medium performance group, and 28\% in the high-performance group.

We determine the angular separation and contrast consistent with a 5$\sigma$ detection by injecting artificial companions, a clone of the primary PSF.\footnote{We find that for Robo-AO data the companion injection method provides a more realistic measure of the detection sensitivity compared to mapping the contrasts consistent with a 5$\sigma$ excursion from the background noise, which results in contrast curves artificially a half-magnitude or more deeper.} For concentric annuli of 0$\farcs$1 width, the detection limit is calculated by repeatedly dimming the artificial companion until the auto-companion detection algorithm (Section \ref{sec:compsearch}) fails to detect it.  This process is subsequently performed at multiple random azimuths within each annulus, and the limiting 5$\sigma$ magnitudes are averaged. For clarity, these average magnitudes for all radii measurements are fitted with functions of the form $a \times sinh(b \times r+c)+d$ (where \textit{r} is the radius from the target star and \textit{a, b, c} and \textit{d} are fitting variables). The limiting contrast curves from observations with Robo-AO at Palomar and Kitt Peak were determined and found to be similar. Typical contrast curves for the three performance groups are shown in Figure$~\ref{fig:contrastcurves}$.

\subsection{Nearby Star Properties}
\label{sec:characterization}

\subsubsection{Contrast Ratios}
\label{sec:contrastratios}

For wide, resolved companions with little PSF overlap, the companion to primary star contrast ratio was determined using aperture photometry on the original images. The aperture radius was cycled in one-pixel increments from 1-5 FWHM for each system, with background measured opposite the primary from the companion (except in the few cases where another object falls near or within this region in the image). Photometric uncertainties are estimated from the standard deviation of the contrast ratios measured for the various aperture sizes.

For close companions, the estimated PSF was used to remove the blended contributions of each star before aperture photometry was performed. The locally optimized PSF subtraction algorithm can attempt to remove the flux from companions using other reference PSFs with excess brightness in those areas. For detection purposes, we use many PSF core sizes for optimization, and the algorithm's ability to remove the companion light is reduced. However, the companion is artificially faint as some flux has still been subtracted. To avoid this, the PSF fit was redone excluding a six-pixel-diameter region around the detected companion. The large PSF regions allow the excess light from the primary star to be removed, while not reducing the brightness of the companion.

\subsubsection{Separation and Position Angles}
\label{sec:separationposangles}

Separation and position angles were determined from the raw pixel positions.  Uncertainties were found using estimated systematic errors due to blending between components. Typical uncertainty in the position for each star was 1-2 pixels. Position angles and the plate scale for observations at Palomar were calculated using a distortion solution produced using Robo-AO measurements for the globular cluster M15.\footnote{S. Hildebrandt (2013, private communication)}

\section{Discoveries} 
\label{sec:Discoveries}
We observed 532 KOIs with Robo-AO, around which we find 94 companions nearby 88 KOIs. 84 of these KOIs with nearby stars have not been previously imaged in high resolution. We find a companion fraction of 16.7$\pm$1.6$\%$ within 4$\farcs$0 of the 532 planetary candidate hosting stars. Cutouts of all multiple star systems are shown in Figures$~\ref{fig:new_cutout_grid1}$ and $~\ref{fig:new_cutout_grid2}$, and measured properties of the systems are detailed in Table$~\ref{tab:newkois_table}$.

In addition, we find 50 additional companions outside 2\farcs5 and within 4\farcs0 around 48 KOIs from 715 targeted KOIs previously observed in Paper I.  Combined with the nearby stars found within 2\farcs5 of the 715 KOIs in Paper I, we detect 103 stars nearby 96 KOIs, for a nearby star fraction rate of 13.4$\pm$1.4$\%$\footnote{Error based on Poissonian statistics \citep{burgasser03}} within 4\farcs0 of a KOI.  Cutouts of the KOIs from Paper I with newly detected nearby stars are shown in Figure$~\ref{fig:paper1_cutouts}$, and measured properties of the systems are detailed in Table$~\ref{tab:paper1_table}$.

\begin{figure}
\centering
\includegraphics[width=0.4\paperwidth]{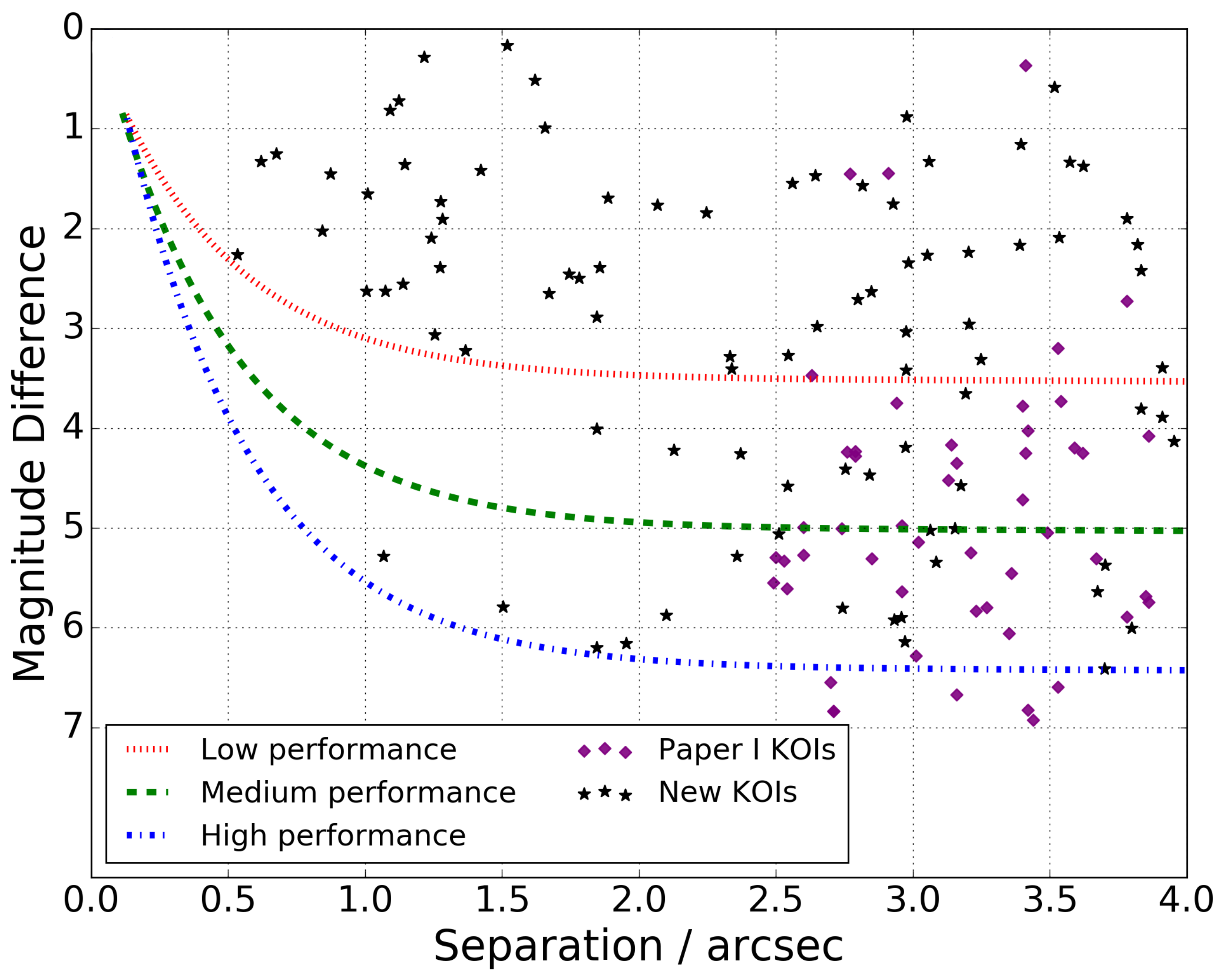}
\caption{Separations and magnitude differences of the detected companions outside 2\farcs5 and within 4\farcs0 from the Paper I targets in black, and from new observations of previously unpublished KOIs in purple. Typical contrasts curves consistent with a 5$\sigma$ detection on low-, medium- and high-performance images are plotted (as described in Section \ref{sec:imageperf}).}
\label{fig:contrastcurves}
\end{figure}

The detected companion separations and contrast ratios of observed nearby stars to KOIs are plotted in Figure$~\ref{fig:contrastcurves}$, along with the calculated 5$\sigma$ detection limits as detailed in Section$~\ref{sec:imageperf}$.

\begin{center}
\begin{table*}
\renewcommand{\arraystretch}{1}
\setlength{\tabcolsep}{4pt}
\caption{Detections of Objects outside 2\farcs5 and within 4\farcs0 of \textit{Kepler} Planet Candidates from Paper I Targets}
\footnotesize
\centering
\begin{tabular}{l c c c c c c c c}
\hline
\hline
\noalign{\vskip 3pt}
KOI & $\rm m_{\textit{Kep}}$ & ObsID & Filter & Det. Significance & Separation & P.A. & Mag. Diff. & Previous \\
& (mag) & & & $\sigma$ & ($\arcsec$) & (deg.) & (mag) &  high res.?  \\
\hline
\noalign{\vskip 3pt}
44 & 13.483 & 2012 Jul 16 & \textit{i} & 8.8 & 3.42$\pm$0.06 & 123$\pm$3 & 4.03$\pm$0.04 & K16\\ 
70 & 12.498 & 2012 Jul 16 & \textit{i} & 20.6 & 3.86$\pm$0.06 & 51$\pm$4 & 5.74$\pm$0.14 & K16\\ 
75 & 10.775 & 2013 Aug 14 & LP600 & 15.9 & 3.53$\pm$0.06 & 124$\pm$4 & 6.6$\pm$0.03 & A12, K16\\ 
99 & 12.960 & 2013 Jul 29 & LP600 & 9.1 & 3.67$\pm$0.06 & 46$\pm$3 & 5.31$\pm$0.03 & K16, L12\\ 
102 & 12.566 & 2013 Oct 25 & LP600 & 15.4 & 2.91$\pm$0.06 & 221$\pm$2 & 1.45$\pm$0.01 & \\ 
107 & 12.702 & 2012 Jul 16 & \textit{i} & 6.9 & 2.6$\pm$0.06 & 273$\pm$3 & 5.27$\pm$0.08 & \\ 
148 & 13.040 & 2012 Jul 17 & \textit{i} & 6.6 & 2.54$\pm$0.06 & 245$\pm$4 & 4.99$\pm$0.06 & K16\\ 
161 & 13.341 & 2012 Jul 18 & LP600 & 18.2 & 2.7$\pm$0.06 & 172$\pm$5 & 6.55$\pm$0.14 & K16\\ 
162 & 13.837 & 2012 Jul 18 & LP600 & 17.9 & 3.23$\pm$0.06 & 0$\pm$4 & 5.83$\pm$0.09 & \\ 
214 & 14.256 & 2012 Jul 18 & LP600 & 13.5 & 3.85$\pm$0.06 & 119$\pm$4 & 5.68$\pm$0.07 & K16 \\ 
220 & 14.236 & 2012 Sep 01 & LP600 & 8.3 & 3.13$\pm$0.06 & 213$\pm$3 & 4.52$\pm$0.04 & \\ 
237 & 14.176 & 2012 Jul 18 & LP600 & 5.5 & 3.16$\pm$0.06 & 208$\pm$4 & 6.67$\pm$0.26 & \\ 
250 & 15.473 & 2012 Aug 03 & LP600 & 3.0 & 3.44$\pm$0.06 & 275$\pm$3 & 6.92$\pm$0.66 & \\ 
263 & 10.821 & 2012 Jul 18 & \textit{i} & 19.0 & 3.34$\pm$0.06 & 267$\pm$2 & 0.59$\pm$0.0 & A12, K16\\ 
268 & 10.560 & 2012 Sep 14 & LP600 & 9.0 & 2.50$\pm$0.06 & 308$\pm$3 & 5.55$\pm$0.01 & A12, K16\\ 
317 & 12.885 & 2012 Jul 28 & \textit{i} & 10.5 & 3.02$\pm$0.06 & 283$\pm$3 & 5.14$\pm$0.06 & \\ 
385 & 13.435 & 2012 Aug 02 & \textit{i} & 9.0 & 3.36$\pm$0.06 & 171$\pm$2 & 5.45$\pm$0.12 & \\ 
465 & 14.188 & 2012 Aug 05 & LP600 & 12.1 & 3.62$\pm$0.06 & 130$\pm$3 & 4.25$\pm$0.07 & L12\\ 
486 & 14.118 & 2012 Aug 05 & LP600 & 13.1 & 3.53$\pm$0.06 & 71$\pm$2 & 3.2$\pm$0.03 & \\ 
509 & 14.883 & 2012 Sep 01 & LP600 & 5.0 & 2.79$\pm$0.06 & 305$\pm$3 & 4.28$\pm$0.14 & \\ 
509 & 14.883 & 2012 Sep 01 & LP600 & 7.6 & 2.94$\pm$0.06 & 55$\pm$2 & 3.75$\pm$0.04 & \\ 
568 & 14.140 & 2012 Aug 05 & LP600 & 8.8 & 3.16$\pm$0.06 & 142$\pm$3 & 4.35$\pm$0.05 & \\
626 & 13.490 & 2012 Aug 03 & \textit{i} & 5.4 & 2.85$\pm$0.06 & 349$\pm$3 & 5.31$\pm$0.04 & L12\\ 
628 & 13.946 & 2012 Aug 03 & \textit{i} & 8.1 & 2.76$\pm$0.06 & 237$\pm$4 & 4.24$\pm$0.06 & L12, F17\\ 
644 & 13.725 & 2012 Aug 04 & \textit{i} & 32.2 & 2.77$\pm$0.06 & 62$\pm$3 & 1.45$\pm$0.01 & L12\\ 
650 & 13.594 & 2012 Aug 04 & \textit{i}  & 21.2 & 2.63$\pm$0.06 & 269$\pm$2 & 3.47$\pm$0.07 & L14, K16\\ 
663 & 13.506 & 2012 Sep 02 & LP600 & 12.5 & 3.21$\pm$0.06 & 61$\pm$3 & 5.8$\pm$0.09 & K16\\ 
685 & 13.949 & 2013 Jul 27 & LP600 & 33.5 & 3.35$\pm$0.06 & 268$\pm$5 & 6.05$\pm$0.12 & L12\\ 
701 & 13.725 & 2012 Aug 05 & \textit{i} & 7.1 & 2.96$\pm$0.06 & 105$\pm$3 & 4.98$\pm$0.06 & K16, F17\\ 
1198 & 15.319 & 2012 Sep 03 & LP600 & 5.0 & 3.11$\pm$0.06 & 98$\pm$4 & 5.25$\pm$0.33 & \\
1279 & 13.749 & 2012 Aug 06 & \textit{i} & 6.1 & 2.74$\pm$0.06 & 134$\pm$3 & 5.0$\pm$0.1 & D14\\ 
1366 & 15.368 & 2012 Sep 04 & LP600 & 10.4 & 3.4$\pm$0.06 & 119$\pm$3 & 4.72$\pm$0.18 & \\ 
1627 & 15.767 & 2012 Sep 04 & LP600 & 27.9 & 3.41$\pm$0.06 & 87$\pm$2 & 0.37$\pm$0.01 & \\ 
1692 & 12.557 & 2012 Aug 29 & \textit{i} & 7.7 & 3.19$\pm$0.06 & 342$\pm$4 & 6.82$\pm$0.13 & W15, K16\\ 
1781 & 12.231 & 2012 Sep 13 & LP600 & 13.6 & 3.4$\pm$0.06 & 331$\pm$3 & 3.78$\pm$0.01 & L12, W15, K16\\ 
1812 & 13.742 & 2012 Aug 29 & \textit{i} & 5.7 & 2.71$\pm$0.06 & 111$\pm$5 & 6.84$\pm$1.58 & L12\\ 
1820 & 13.530 & 2012 Sep 13 & LP600 & 37.9 & 3.78$\pm$0.06 & 180$\pm$4 & 5.89$\pm$0.08 & \\ 
1845 & 14.438 & 2013 Oct 25 & LP600 & 19.1 & 3.04$\pm$0.06 & 347$\pm$3 & 4.59$\pm$0.09 & \\ 
1884 & 15.462 & 2012 Sep 13 & LP600 & 5.8 & 2.54$\pm$0.06 & 328$\pm$4 & 5.61$\pm$0.47 & B16\\ 
1922 & 15.356 & 2012 Sep 13 & LP600 & 24.5 & 3.78$\pm$0.06 & 195$\pm$2 & 2.73$\pm$0.03 & \\ 
2022 & 14.746 & 2012 Sep 13 & LP600 & 10.4 & 3.14$\pm$0.06 & 71$\pm$3 & 4.16$\pm$0.13 & \\ 
2022 & 14.746 & 2012 Sep 13 & LP600 & 7.8 & 2.5$\pm$0.06 & 152$\pm$3 & 5.3$\pm$0.26 & \\ 
2025 & 13.781 & 2012 Sep 13 & LP600 & 15.9 & 3.49$\pm$0.06 & 191$\pm$4 & 5.05$\pm$0.02 & \\ 
2105 & 13.862 & 2012 Oct 06 & LP600 & 7.6 & 3.01$\pm$0.06 & 314$\pm$3 & 5.28$\pm$0.11 & \\ 
2169 & 12.404 & 2012 Aug 31 & \textit{i} & 30.1 & 3.59$\pm$0.06 & 66$\pm$3 & 4.2$\pm$0.01 & W15, K16\\ 
2222 & 12.963 & 2012 Aug 31 & \textit{i} & 6.1 & 2.53$\pm$0.06 & 333$\pm$3 & 5.33$\pm$0.1 & \\ 
2287 & 12.485 & 2012 Aug 31 & \textit{i} & 9.8 & 2.96$\pm$0.06 & 11$\pm$4 & 5.64$\pm$0.09 & K16\\ 
2547 & 14.169 & 2012 Oct 06 & LP600 & 13.7 & 2.79$\pm$0.06 & 151$\pm$3 & 4.23$\pm$0.02 & \\ 
2556 & 14.050 & 2012 Oct 06 & LP600 & 17.7 & 3.86$\pm$0.06 & 238$\pm$2 & 4.08$\pm$0.05 & \\ 
2582 & 13.628 & 2012 Aug 31 & \textit{i} & 12.3 & 3.41$\pm$0.06 & 223$\pm$2 & 4.25$\pm$0.05 & \\ 
2641 & 13.845 & 2012 Oct 06 & LP600 & 17.9 & 3.54$\pm$0.06 & 0$\pm$2 & 3.73$\pm$0.01 & \\ 

\hline
\end{tabular}
\small
\label{tab:paper1_table}
\begin{flushleft}
Notes. --- References for previous high-resolution observations are denoted using the following codes: \citealt{adams12} (A12),  \citealt{lillo12} (L12), \citealt{dressing14} (D14), \citealt{kraus16} (K16), \citealt{wang15a} (W16), \citealt{baranec16} (B16), \citealt{furlan16} (F17)
\end{flushleft}
\end{table*}
\end{center}

\subsection{Comparison to Other Surveys}
\label{sec:othersurveys}

Some of the KOIs with observations presented in this paper have been previously observed in other surveys.  In this section, we compare our nearby star detections and non-detections with the observations from other telescopes.

\citet{lillo12} and \citet{lillo14} observed 98 and 174 KOIs, respectively, using the AstraLux Lucky Imaging system on the 2.2m telescope at the Calar Alto Observatory.  The nearby stars to KOI-99, 465, 626, 628, 644, 685, 1781, and 1812, all from the Paper I target list, were previously detected by them, as well as KOI-3805 from the targets observed with Robo-AO presented in this work.  We did not detect the nearby star to KOI-238 from \citet{lillo12} with $\Delta$m$_{J}$=4.38; this star may be significantly fainter in the visible leading to our non-detection.  We also do not detect the nearby stars from \citet{lillo14} to KOI-1230 ($\Delta$m$_{i'}$=9.11) and KOI-2324 ($\Delta$m$_{i'}$=6.12), which are outside our detection sensitivity.

\citet{wang15a} observed 84 KOIs using the PHARO and NIRC2 instruments at Palomar and Keck, respectively.  We observe nearby stars to KOI-1692, 1781, and 2169 which they previously detected, but do not observe their detected nearby stars to KOI-344 ($\Delta$m$_{J}$=5.52), KOI-1353 ($\Delta$m$_{J}$=4.87), KOI-5515 ($\Delta$m$_{J}$=4.10 and $\Delta$m$_{J}$=5.40). The NIR photometry of the nearby stars to KOI-344 and 5515 suggest they are later spectral types than the target star and may be faint in the visible.  The apparent visual magnitude of the nearby star to KOI-1353 is not known, and it may be too faint for detection in this survey.

\citet{adams12} and \citet{adams13} observed 87 and 13 KOIs with the instruments ARIES and PHARO on the MMT and Palomar telescopes, respectively. We observe the nearby stars to KOI-75, 263, and 268 at separations greater than 2\farcs5 which they had previously detected.  They detect nearby stars to KOI-10, 18, 113, 1316, all with $\Delta$m$_{J}>$6.0, which we did not detect as they likely have contrast ratios in the visible outside our detection sensitivity.

Observing 87 KOIs with ARIES at the MMT, \citet{dressing14} previously detected a nearby star to KOI-1279. In addition, they detected nearby stars to KOI-720 ($\Delta$m$_{J}$=5.50) and KOI-2331 ($\Delta$m$_{J}$=4.55) which we did not detect.  NIR photometry suggests these nearby stars are redder than the target star, making them too faint for our survey to detect.

\citet{gilliland15} detected a nearby star to KOI-2650 ($\Delta$m$_{755W}$=7.55) using the \textit{Hubble Space Telescope}, too faint for detection in this survey.

\citet{kraus16} observed 382 KOIs with AO on the Keck-II telescope.  We observe nearby stars they previously detected to KOI-44, 70, 75, 99, 102, 148, 161, 214, 263, 268, 663, 701, 1692, 1781, 2169, and 2287.  They detect faint ($\Delta$m$_{K}>$6.0) stars, below our detection sensitivity, at separations outside 2\farcs5 to Paper I targets: KOI-2, 41, 84, 85, 103, 105, 144, 152, 157, 177, 254, 261, 269, 372, 571, 701, 886, 899, 947, 1146, 1230, 1241, 1316, 1408, 1589, 1615, 1618, 1738, 1843, 2158, 2332, and 2593.  They also detect faint nearby stars to KOI-72, 2650, and 2792 that were also observed with Robo-AO and presented in this work.  These high-contrast stars are outside of our detection sensitivity.

\citet{furlan16} observed 253, 317, and 310 unique KOI host stars at Keck, Palomar, and Lick Observatory, respectively.  We observe nearby stars previously detected by \citet{furlan16} to KOI-628, 701, 959, 980, 1614, 3156, 5475, 6600, 7032, 7455, and 7470.  

We also detect the nearby star to KOI-1884 discovered with Keck-AO in Paper II.

In summary, we detect every star discovered by other surveys near the observed KOIs that have separations and contrasts within our detection sensitivities. Many previous surveys were performed with large-aperture telescopes in the NIR, sensitive to faint and red companions, which we do not detect in this survey. If there is a true planet hosted by the primary star in the system, the flux from these nearby stars will have a negligible effect on the visible \textit{Kepler} light curves.

\section{Discussion}
\label{sec:Discussion}

In this section, we present comprehensive statistics for the Robo-AO \textit{Kepler} survey, as well as discuss the impact of the detected nearby stars on the planetary candidate properties.

\subsection{Robo-AO KOI Survey Cumulative Statistics}

\begin{figure}
\centering
\includegraphics[width=0.4\paperwidth]{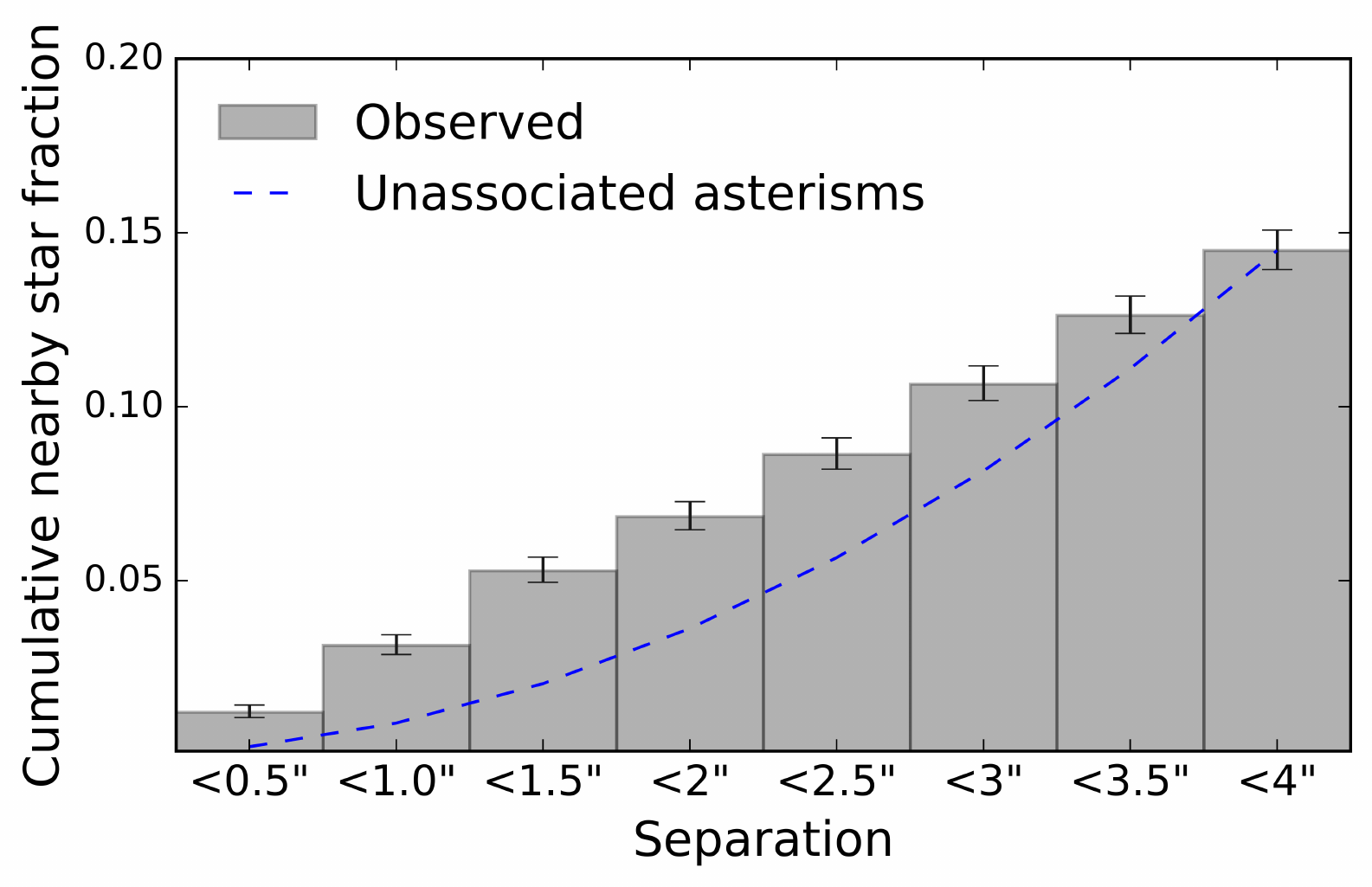}
\caption{The nearby star fraction rate as a function of separation from 3857 observations of planetary candidates in the Robo-AO KOI survey. The dashed line represents a cumulative distribution that scales with the area that would be expected from non-physically-associated companions.}
\label{fig:rates}
\end{figure}

\begin{figure*}
\centering
\includegraphics[width=500pt]{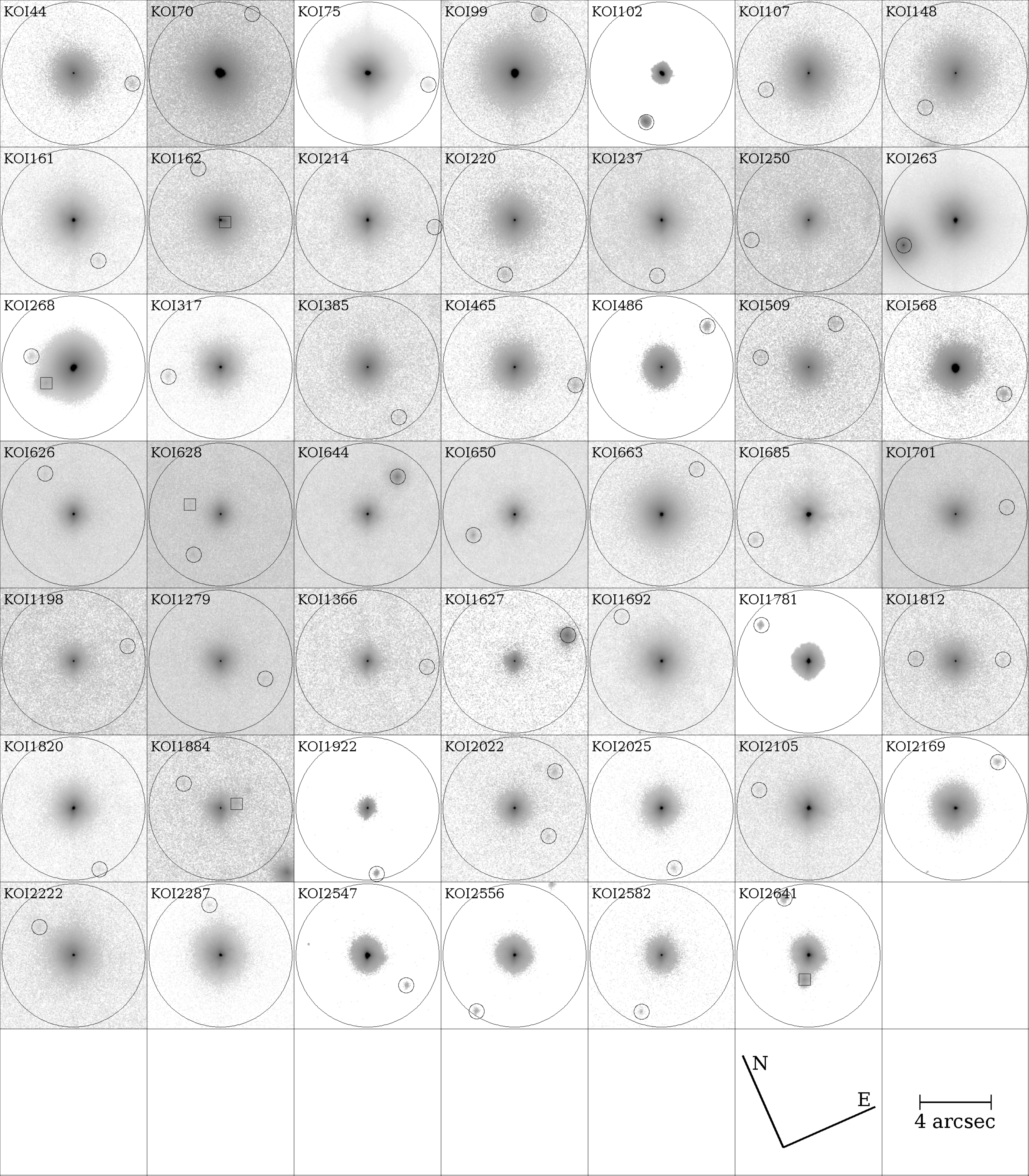}
\caption{Color inverted, normalized log-scale cutouts of 48 multiple KOI systems with separations outside 2$\farcs$5 and within 4$\arcsec$ resolved with Robo-AO at Palomar from the Paper I target list.  The angular scale and orientation are similar for each cutout.  The smaller circles are centered on the detected nearby star, and the larger circle is the limit of the survey's 4\arcsec separation range. Squares are centered on companions with separations less than 2$\farcs$5 found in Paper I from Robo-AO at Palomar.}
\label{fig:paper1_cutouts}
\end{figure*}

\begin{figure*}
\centering
\includegraphics[width=500pt]{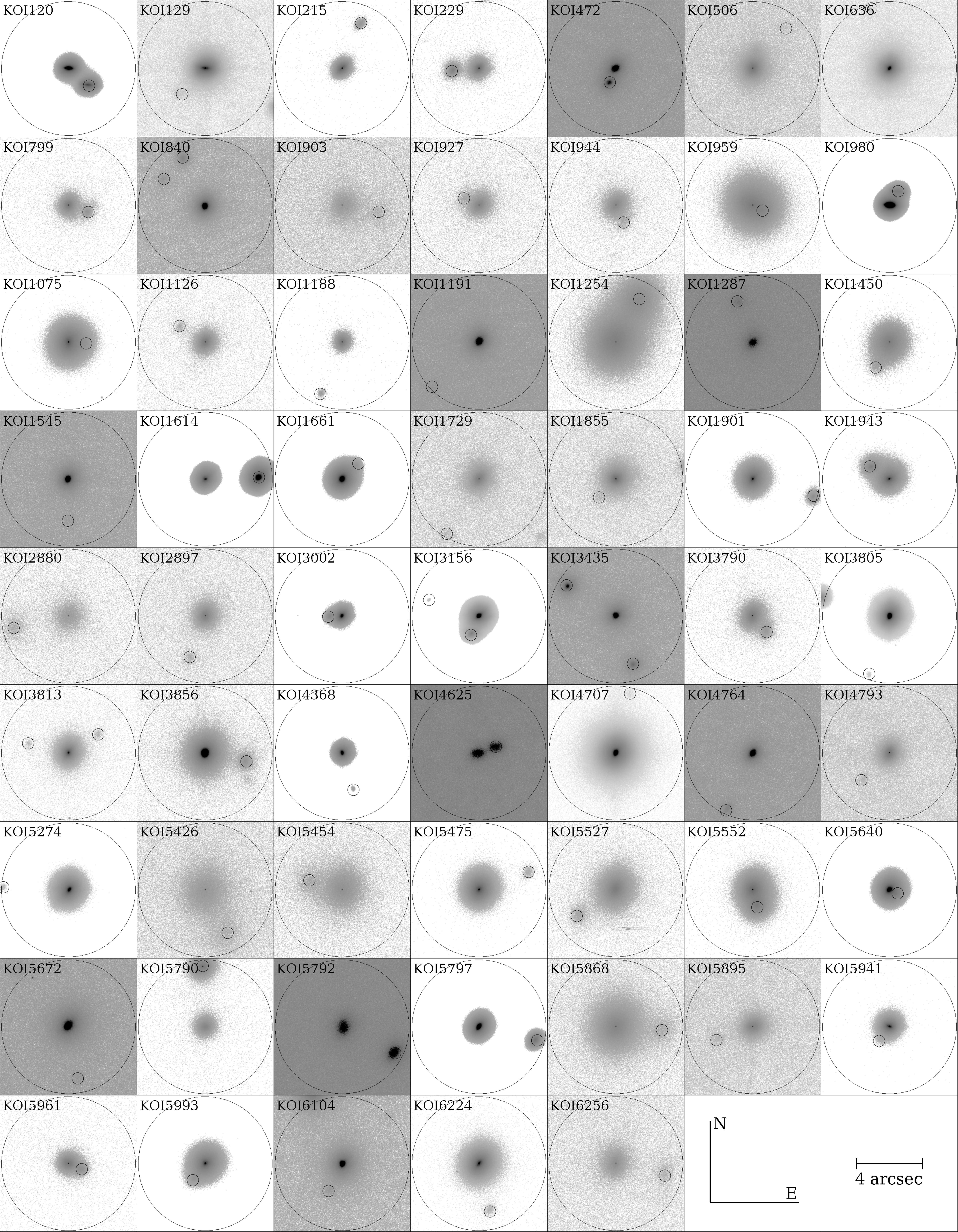}
\caption{Color inverted, normalized log-scale cutouts of 61 multiple KOI systems [KOI-120 to KOI-6256] resolved with Robo-AO from Kitt Peak.  The angular scale and orientation are similar for each cutout.  The smaller circles are centered on the detected nearby star, and the larger circle is the limit of the survey's 4\arcsec separation range.}
\label{fig:new_cutout_grid1}
\end{figure*}

\begin{figure*}
\centering
\includegraphics[width=500pt]{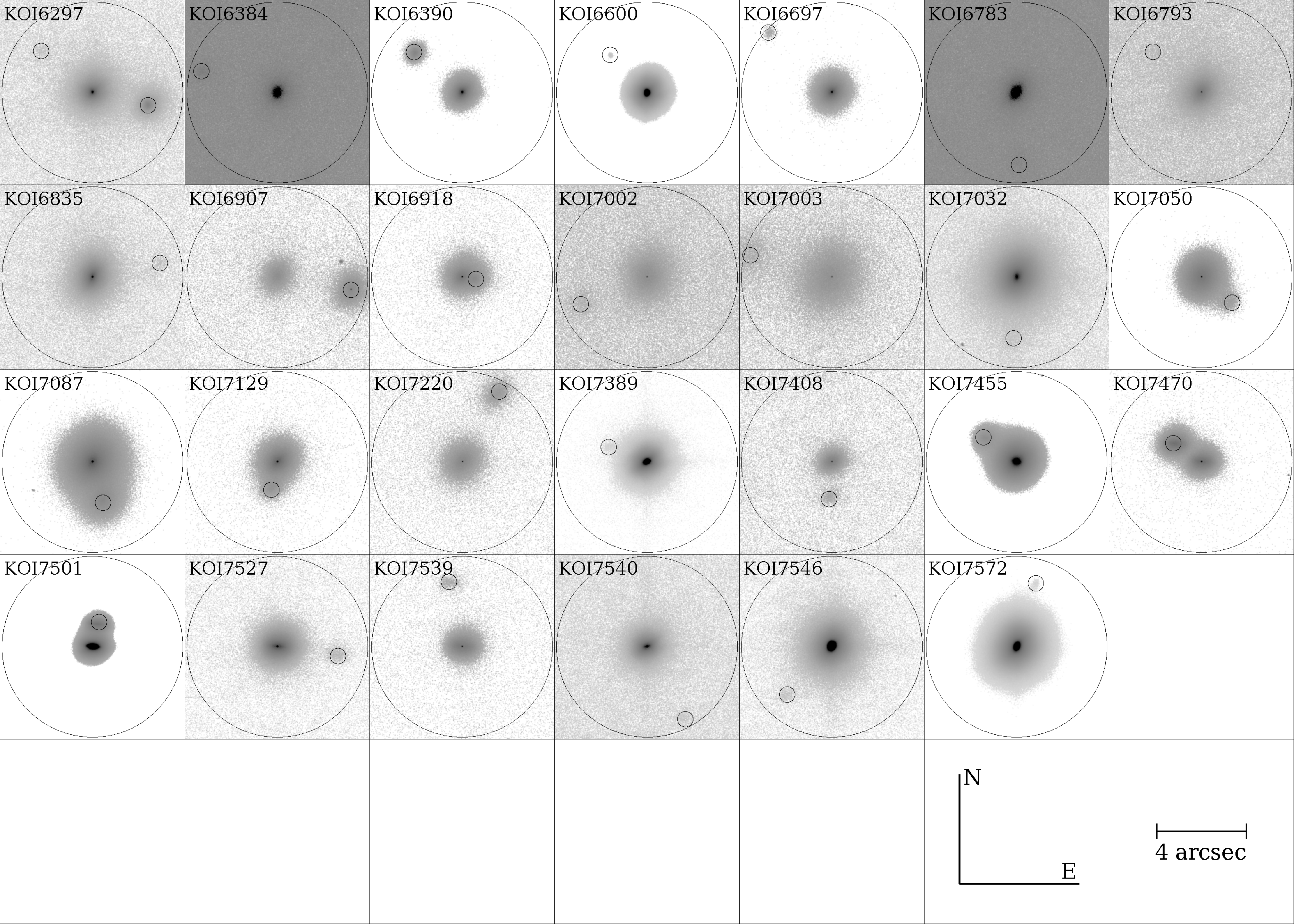}
\caption{Color inverted, normalized log-scale cutouts of 27 multiple KOI systems [KOI-6297 to KOI-7572] resolved with Robo-AO from Kitt Peak.  The angular scale and orientation are similar for each cutout.  The smaller circles are centered on the detected nearby star, and the larger circle is the limit of the survey's 4\arcsec separation range.}
\label{fig:new_cutout_grid2}
\end{figure*}

The Robo-AO KOI survey has observed 3857 KOIs in Paper I, II, III, and this work. We find 610 nearby stars around 559 planetary candidate hosts in the combined survey dataset, implying a nearby star fraction rate of 14.5$\pm$0.6\% within the Robo-AO detectability range (separations between $\sim$0$\farcs$15 and 4$\farcs$0 and $\Delta$m$\le$6). We also find within 4\farcs0 separation, a triple star fraction of 1.2$\pm$0.2\% and a quadruple star fraction of $0.08^{+0.06}_{-0.03}\%$. The nearby star fraction rate as a function of separation from the host star for the survey to date is listed in Table$~\ref{tab:rates}$ and plotted in Figure$~\ref{fig:rates}$. The nearby star fraction increases linearly with separation from the host star. If all nearby stars were unbound, we would expect the rate to increase with the area enclosed. This suggests that a significant fraction of the nearby stars may be bound to the host star. It should be noted that this analysis does not account for the detection sensitivity of Robo-AO at varying separations. It is expected, however, that most nearby stars at separations $<$1\arcsec are likely bound \citep{horch14}. We will assess the probability of association of individual systems in future papers in this survey.

The properties of planetary systems in binary star systems may be impacted due to perturbations from the secondary star. We show in Table$~\ref{tab:planettyperates}$ the nearby star fraction for different planet types based on their similarity in radius to a solar system planet. We find that the nearby star rates for all four planet types are within 2$\sigma$ of the total rate for the entire survey. The largest outlier rate is for the Jupiter or gas giant planets, which are known to have a large false positive fraction \citep{santerne13}, caused by the potential of background eclipsing binaries to mimic their deep transits. \citet{wang14} also find a high stellar multiplicity rate for hot Jupiters, and direct imaging surveys find that gas giants have a high rate of bound stellar companions \citep{ngo16, evans16}. It is also possible that the high nearby star rate may be due to orbital migration caused by a bound secondary star which drives gas giants to low period orbits more easily detectable by \textit{Kepler}.  In Paper III, we found a significant increase in the nearby star rate for low-period giant planets, possibly caused by orbital migration due to the secondary star \citep{fabrycky07}, although the significance of this effect may be small \citep{naox12, petrovich15}. These migrations may also cause planet scattering, differentially ejecting smaller planets from the system \citep{rasio96, wang15a}.

\begin{table}
\begin{center}

\renewcommand{\arraystretch}{1.3}
\setlength{\tabcolsep}{4pt}
\caption{\label{tab:rates}Robo-AO KOI Survey Cumulative Nearby Star Fraction Rates}
\begin{tabular}{ccc}
\hline
\hline
\noalign{\vskip 3pt}
\colhead{Separation} & \colhead{Systems with nearby stars} & \colhead{Nearby star rate\footnote{Error based on Poissonian statistics \citep{burgasser03}}} \\
\colhead{(\arcsec)} &  & \colhead{(\%)} \\
\hline
\noalign{\vskip 3pt}
$<$0.5 & 47 & $1.2^{+0.20}_{-0.15}$ \\
$<$1.0 & 121 & $3.1^{+0.30}_{-0.26}$ \\
$<$1.5 & 204 & $5.3^{+0.38}_{-0.33}$ \\
$<$2.0 & 264 & $6.8^{+0.43}_{-0.38}$ \\
$<$2.5 & 333 & $8.6^{+0.47}_{-0.43}$ \\
$<$3.0 & 411 & $10.7^{+0.52}_{-0.48}$ \\
$<$3.5 & 487 & $12.6^{+0.55}_{-0.52}$ \\
$<$4.0 & 559 & $14.5^{+0.59}_{-0.55}$ \\

\hline
\end{tabular}
\small
\end{center}
\end{table}

\begin{table}
\begin{center}

\renewcommand{\arraystretch}{1.3}
\setlength{\tabcolsep}{1pt}
\caption{\label{tab:planettyperates}Nearby Star Fraction Rates By Planet Candidate Type}
\begin{tabular}{ccccc}
\hline
\hline
\noalign{\vskip 3pt}
\colhead{Planet candidate} & \colhead{Planetary radius} & \colhead{Systems with} & \colhead{Total} & \colhead{Nearby star} \\
\colhead{type} & \colhead{range} & \colhead{nearby stars} & \colhead{systems} & \colhead{rate} \\
\hline
\noalign{\vskip 3pt}
Earths & R$_{p}<$1.6$R_{\oplus}$ & 241 & 1480 & $16.3\pm1.0\%$ \\
Neptunes & 1.6$R_{\oplus}<R_{p}<3.9R_{\oplus}$ & 268 & 2058 & $13.0\pm0.8\%$ \\
Saturns & 3.9$R_{\oplus}<R_{p}<9R_{\oplus}$ & 46 & 338 & $13.6\pm2.0\%$ \\
Jupiters & 9$R_{\oplus}<R_{p}$ & 47 & 247 & $19.0\pm2.8\%$ \\
\hline
\end{tabular}
\small
\end{center}
\end{table}

\subsection{Implications for \textit{Kepler} Planet Candidates}
\label{sec:implications}

A nearby star in the same photometric aperture as the target star will dilute the observed transit depth, resulting in underestimated radius estimates.  We re-derive the estimated planetary radii for the 814 planetary candidates around the 559 KOIs with detected nearby stars in the Robo-AO \textit{Kepler} survey for two scenarios: 1) the planet orbits the target star; and 2) the planet orbits the secondary star which is bound to the primary star\footnote{The third scenario, in which the secondary star is unbound to the primary star, is unconstrained without color information. Future papers in this survey will address the implications on the radius of \textit{Kepler} planetary candidates in this scenario.}. For the first case, we use the relation from Paper I to correct for the transit dilution,
\begin{equation}
R_{p,A}=R_{p,0}\sqrt{\frac{1}{F_{A}}}
\end{equation}
where R$_{p,A}$ is the corrected radius of the planet orbiting the primary star, R$_{p,0}$ is the original planetary radius estimate based on the diluted transit signal, and F$_{A}$ is the fraction of flux within the aperture from the primary star.  For the case where the planet candidate is bound to the secondary star, we use the relation
\begin{equation}
R_{p,B}=R_{p,0}\frac{R_{B}}{R_{A}}\sqrt{\frac{1}{F_{B}}}
\end{equation}
where R$_{p,B}$ is the corrected radius of the planet orbiting the secondary star bound to the primary star, R$_{B}$ and R$_{A}$ are the stellar radii of the secondary and primary star, respectively, and F$_{B}$ is the fraction of flux within the aperture from the secondary star.

We use the stellar radius estimates from \citet{dr25} for the primary stars. The radii of secondary companions in the scenario where they are bound to the target star were estimated using the observed contrast ratio in the \textit{Kepler} band (approximated using the LP600 bandpass) and finding the radius of an appropriately fainter star within the Dartmouth stellar models \citep{dotter08}. The fluxes of all observed sources within the \textit{Kepler} aperture were summed to estimate the transit dilution. The revised planetary radius estimates are detailed in Table$~\ref{tab:radii}$.

The original derived planetary candidate radius estimates are corrected for dilution only from nearby stars resolved in the KIC \citep{coughlin15}. We find that four of the nearby stars detected in our survey appear in the KIC (companions to KOIs 263, 521, 1614, and 5790). We therefore do not revise the radius estimates for the planetary candidates in these four systems and they are not included in the the following analysis.

Of the 814 planetary candidates with nearby stars detected in this survey, approximately 29$\%$ have a corrected planetary radius at least 10$\%$ larger than the original planetary radius estimate, assuming the planet candidate orbits the primary star. If instead the planet candidate orbits the secondary star which is bound to the primary star, almost every (99$\%$) planetary candidate has a corrected radius greater than 10$\%$ larger than the original radius estimate. 

If all planet candidates orbit the primary star, the original planetary radii derived from the \textit{Kepler} light curves are underestimated by a factor of 1.08, on average. If all planet candidates instead orbit the secondary star which is bound to the primary, the corrected planetary radius estimates are on average a factor of 3.29 larger than those originally derived. The more realistic scenario is if we assume that the planet candidates are equally likely to be orbiting the primary or secondary stars. In this case, the radius estimates for the planetary candidate in systems with nearby stars will increase by a factor of 2.18 on average. This is significantly higher than the radius correction factor of 1.6 found by \citet{ciardi15} and 1.64 found by \citet{hirsch17}. \citeauthor{hirsch17} used planetary occurrence rates \citep{howard12} to estimate the fraction of planets orbiting the primary and secondary star for known bound systems. It is unclear, however, if this approach results in a more accurate planetary correction factor estimate, because, as they note, the planet occurrence rates in binary systems is not well understood. Indeed, we found evidence in Paper III that binary stars significantly affect the properties of planetary systems, driving migration of large planets to low-period orbits.

\begin{table}
\begin{center}

\renewcommand{\arraystretch}{1.3}
\setlength{\tabcolsep}{4pt}
\caption{\label{tab:notrocky}Planetary Candidates Likely Not Rocky Due to Nearby Stars}
\begin{tabular}{cccccc}
\hline
\hline
\noalign{\vskip 3pt}
\colhead{Object} & \colhead{Sep.} & \colhead{$\Delta$m} &  \colhead{R$_{p,0}$\footnote{Original planetary radius estimate, from NASA Exoplanet Archive.}} &  \colhead{R$_{p,prim.}$\footnote{Estimated planetary radius in the scenario where the planet orbits the target star.}} & \colhead{R$_{p,sec.}$\footnote{Estimated planetary radius in the scenario where the planet orbits the secondary star, which is physically bound to the target star.}} \\
\colhead{} & \colhead{(\arcsec)} & \colhead{(mags)} & \colhead{(R$_{\Earth}$)} & \colhead{(R$_{\Earth}$)}  & \colhead{(R$_{\Earth}$)} \\
\hline
\noalign{\vskip 3pt}
0284.02 & 0.96 & 0.45 & 1.40 & 1.80 & 2.0 \\ 
0284.04 & 0.96 & 0.45 & 1.41 & 1.82 & 2.1 \\ 
0298.01 & 2.01 & 0.58 & 1.50 & 1.89 & 2.3 \\ 
1214.01 & 0.33 & 1.21 & 1.44 & 1.66 & 2.4 \\ 
1630.01 & 1.77 & 0.91 & 1.40 & 1.68 & 2.3 \\ 
1700.01 & 0.29 & 1.07 & 1.54 & 1.8 & 2.6 \\ 
1973.01 & 0.79 & 1.69 & 1.49 & 1.64 & 3.4 \\ 
2163.03 & 0.77 & 0.04 & 1.59 & 2.23 & 2.2 \\ 
2377.01 & 2.09 & 1.25 & 1.55 & 1.78 & 2.7 \\ 
2486.01 & 0.24 & 0.49 & 1.42 & 1.82 & 2.0 \\ 
2551.01 & 2.69 & 1.93 & 1.53 & 1.65 & 3.3 \\ 
2580.01 & 0.60 & 0.86 & 1.59 & 1.92 & 2.5 \\ 
2598.01 & 1.09 & 0.37 & 1.35 & 1.77 & 2.0 \\ 
2711.02 & 0.52 & 0.12 & 1.43 & 1.97 & 2.0 \\ 
2851.02 & 0.39 & 0.45 & 1.50 & 1.93 & 2.2 \\ 
2896.02 & 0.96 & 0.38 & 1.57 & 2.05 & 2.3 \\ 
3029.02 & 0.28 & 0.68 & 1.35 & 1.67 & 2.1 \\ 
3112.01 & 1.87 & 0.49 & 1.41 & 1.8 & 2.1 \\ 
3120.01 & 1.14 & 0.87 & 1.43 & 1.72 & 2.2 \\ 
3214.01 & 0.49 & 0.73 & 1.53 & 1.88 & 2.2 \\ 
3214.02 & 0.49 & 0.73 & 1.35 & 1.66 & 2.0 \\ 
3435.01 & 3.06 & 1.33 & 1.58 & 1.8 & 2.8 \\ 
3435.01 & 3.52 & 0.58 & 1.58 & 1.99 & 2.3 \\ 
3928.01 & 2.96 & 1.21 & 1.45 & 1.67 & 2.3 \\ 
4021.01 & 1.92 & 0.52 & 1.53 & 1.95 & 2.4 \\ 
4323.01 & 1.12 & 2.22 & 1.59 & 1.69 & 3.2 \\ 
4331.01 & 0.45 & 0.25 & 1.45 & 1.94 & 2.1 \\ 
4463.01 & 2.45 & 0.01 & 1.52 & 2.14 & 2.1 \\ 
4759.01 & 0.67 & 2.12 & 1.54 & 1.65 & 3.3 \\ 
4823.01 & 1.40 & 0.59 & 1.51 & 1.9 & 2.3 \\ 
5274.01 & 3.95 & 4.13 & 1.59 & 1.61 & 5.7 \\ 
5762.01 & 0.23 & 0.65 & 1.37 & 1.71 & 2.2 \\ 
6475.01 & 1.31 & 0.5 & 1.54 & 1.97 & 2.3 \\ 
6482.01 & 0.52 & 0.58 & 1.53 & 1.93 & 2.4 \\ 
6907.01 & 3.35 & -0.36 & 1.14 & 1.76 & 1.6 \\ 
\hline
\end{tabular}
\small
\end{center}
\end{table}

The large number of unbound background stars likely inflates our estimates of the planetary correction radius factor. These stars are often much fainter than the primary star and the assumption that each star is equally likely to host the planet results in a large number of gas giant planets, which are inherently rare compared to terrestrial planets \citep{howard12}. Simulations from galactic stellar models suggest that the majority of nearby stars to KOIs at separations larger than 1\arcsec are likely unbound \citep{horch14}, a conclusion borne out by observations \citep{atkinson16, hirsch17}. If we limit our survey to just those likely bound nearby stars within 1\arcsec, we find radius correction factors of 1.18, 1.88, and 1.54 for the scenarios where all planets orbit the primary star, all planets orbit a bound secondary star, and all planets are equally likely to orbit either star, respectively. The radius correction factors found for the set of likely bound stars is in agreement with that found by \citet{hirsch17}, and is our recommended estimate for the true radius correction factor for \textit{Kepler} planetary candidates with detected nearby stars. We will quantify the probability of association for every detected nearby star in future papers in this survey, allowing us to better remove unbound background stars from our sample and revisit this discussion.

Lastly, using the original estimates for planetary radius and the planetary radius ranges listed in Table$~\ref{tab:planettyperates}$, we find the radius correction factor for systems with nearby stars within 4\arcsec (1\arcsec) for Earth-sized planets is 2.30 (1.54), for Neptune-sized planets is 2.25 (1.59), for Saturn-sized planets is 1.95 (1.67), and for Jupiter-sized planets is 1.88 (1.38), if we assume that each nearby star is bound and the planetary candidate is equally likely to orbit the primary or secondary star. Under these same assumptions, we estimate that approximately 140 previously believed rocky planet candidates (R$_{p,0}$$<$1.6$R_{\oplus}$), or 9\% of the 1480 rocky planet candidates discovered by \textit{Kepler}, have corrected radii larger than the rocky planet cutoff at 1.6$R_{\oplus}$ as described in \citet{rogers15} due to nearby stars within 4\arcsec. These 140 planetary candidates are therefore likely not rocky due to incorrect identification of the planetary host star and photometric contamination from nearby stars.

We also find 35 rocky planet candidates that, due to the presence of a previously undetected nearby star, are now likely not rocky if either orbiting the primary or secondary stars. We highlight these planetary candidates in Table$~\ref{tab:notrocky}$.

\begin{table}
\begin{center}

\renewcommand{\arraystretch}{1.3}
\setlength{\tabcolsep}{4pt}
\caption{\label{tab:habitable}Implications on Derived Radius of Potentially Rocky, Habitable Zone Planets}
\begin{tabular}{cccccc}
\hline
\hline
\noalign{\vskip 3pt}
\colhead{Object} & \colhead{Equil. Temp.\footnote{Estimated planetary equilibrium temperature, from NASA Exoplanet Archive.}} & \colhead{R$_{p,0}$\footnote{Original planetary radius estimate, from NASA Exoplanet Archive.}} &  \colhead{R$_{p,prim.}$\footnote{Estimated planetary radius in the scenario where the planet orbits the target star.}} & \colhead{R$_{p,sec.}$\footnote{Estimated planetary radius in the scenario where the planet orbits the secondary star, which is physically bound to the target star.}} \\
\colhead{} & (K) & \colhead{(R$_{\Earth}$)} & \colhead{(R$_{\Earth}$)}  & \colhead{(R$_{\Earth}$)} \\
\hline
\noalign{\vskip 3pt}
0701.03 & 269 & 1.72 & 1.73 & 10.94  \\ 
0701.04 & 207 & 1.43 & 1.44 & 9.1  \\ 
7470.01 & 225 & 1.9 & 2.59 & 2.67  \\ 
\hline
\end{tabular}
\small
\end{center}
\end{table}

\subsection{Rocky, Habitable Zone Candidates}

A primary objective of the \textit{Kepler} mission was to estimate $\eta_{\Earth}$, the occurrence rate of Earth-like planets orbiting in the habitable zone.  Contamination from nearby stars has a significant effect on the derived planetary radii. Planetary radii based on \textit{Kepler} light curves alone are underestimated by a factor of approximately 1.5 on average, as discussed in Section$~\ref{sec:implications}$. The impact of nearby stars must, therefore, be taken into account to estimate precisely what planets are terrestrial.  While the exact requirements for habitability remain unclear \citep{kasting93, selis07, seager13, zsom13}, it is believed that the equilibrium temperature of the planet must allow the presence of liquid water. To be Earth-like, a planet must also be rocky: \citet{rogers15} show that the transition between ``rocky'' and ``non-rocky'' occurs rather sharply at R$_{P}$=1.6$R_{\oplus}$.

We searched for potentially rocky planets, with estimated radii less than 2$\sigma$ away from the rocky planet cutoff of 1.6$R_{\oplus}$, residing in the habitable zone (estimated planetary equilibrium temperature $\le$370 K) within the set of systems with newly discovered nearby stars. We find three such planetary candidates, detailed in full in Table$~\ref{tab:radii}$ and highlighted in Table$~\ref{tab:habitable}$.

The two confirmed planets, KOI-701.03 and 701.04 (Kepler-62e and Kepler-62f, respectively), both reside in the habitable zone if orbiting the primary star.  If instead, either one orbits the faint secondary star and that star is bound to the primary, the estimated radii of each would be much larger and it would be unlikely that they would be rocky in composition. This planet has been thoroughly vetted by \citet{borucki13}, who concluded that the two planets are indeed rocky and orbit in the habitable zone.

KOI-7470.01 has an original radius estimate of 1.9$R_{\oplus}$, near the rocky planet cutoff, and an estimated equilibrium temperature of 225 K.  The undiluted radius estimate for the scenario where the planetary candidate orbits the primary is 2.59$R_{\oplus}$, making it very improbable that the planet is rocky.  Likewise, if the planetary candidate instead orbits the bound secondary star, it would again be unlikely to be rocky, with a planetary radius estimate of 2.70$R_{\oplus}$.

\section{Conclusion}
\label{sec:conclusion}

Combining the data sets from the complete Robo-AO KOI survey, we found 610 nearby stars around 559 planetary candidate hosts, from a target list of 3857 KOIs, implying a nearby star fraction rate of 14.5\%$\pm$0.6\% within the Robo-AO detectability range (separations between $\sim$0$\farcs$15 and 4$\farcs$0 and $\Delta$m$\le$6). We found a nearby star fraction for Earth-sized planets of $16.3\pm1.0\%$, for Neptune-sized planets of $13.0\pm0.8\%$, for Saturn-sized planets of $13.6\pm2.0\%$, and for Jupiter-sized planets of $19.0\pm2.8\%$. We derived the corrected planetary radius for every planetary candidate with nearby stars in this survey. We found that planets in systems with likely bound nearby stars have underestimated radii by a factor of 1.54, if we assume each planet is equally likely to orbit the primary or secondary star. We found that 35 of the previously believed rocky planet candidates detected by \textit{Kepler} are likely not rocky due to the presence of a nearby star.

We have also recently made the results of our survey available at a survey website.\footnote{\url{http://roboaokepler.org/}}

In future papers in this analysis, we will use the nearly 4000 high-resolution images of planetary candidate hosts to search for insight into the how binary stars impact planetary formation and evolution. In 2017, we began a campaign to characterize the detected nearby stars to planetary candidate hosts with multi-band photometry. This study will allow the probability of association between stars in each system to be quantified. We are also studying the potential of AO transit observations to detect the source of the transit signal in multiple star systems. While the transit of many \textit{Kepler} planets will likely be too shallow to detect with Robo-AO, we could detect deeper transits from background eclipsing binaries that, when blended with the bright primary stars, are the source of false positive planetary transit signals. 

A second generation Robo-AO instrument on the University of Hawai`i 2.2-m telescope on Maunakea \citep{Robo-AO2} is being built.  The Kitt Peak and Maunakea systems will together image up to $\sim$500 objects per night and have access to three-quarters of the sky over the course of a year.  A southern analog to Robo-AO mounted on the Southern Astrophysical Research Telescope (SOAR) at CTIO and capable of twice HST resolution imaging, is also in development \citep{robosoar}.  With unmatched efficiency, Robo-AO and its lineage of instruments are uniquely able to perform high-acuity imaging of the hundreds of K2 \citep{K2} planetary candidates, ground-based transit surveys such as MEarth \citep{mearth}, KELT \citep{kelt1, kelt2}, HATNet \citep{hatnet}, SuperWASP \citep{superwasp}, NGTS \citep{ngts}, XO \citep{xo}, and the Evryscope \citep{evryscope}, as well as the thousands of expected exoplanet hosts discovered by the forthcoming NASA Transiting Exoplanet Survey Satellite \citep[TESS,][]{TESS} and ESA PLAnetary Transits and Oscillations of stars 2.0 \citep[PLATO,][]{PLATO} missions.

\section*{Acknowledgements}
We thank the anonymous referee for careful analysis and useful comments on the manuscript.

This research is supported by the NASA Exoplanets Research Program, grant $\#$NNX 15AC91G. C.Z. and W.H. acknowledge support from the North Carolina Space Grant consortium.  C.B. acknowledges support from the Alfred P. Sloan Foundation. 

The Robo-AO team thanks NSF and NOAO for making the Kitt Peak 2.1-m telescope available. We thank the observatory staff at Kitt Peak for their efforts to assist Robo-AO KP operations. Robo-AO KP is a partnership between the California Institute of Technology, the University of Hawai‘i, the University of North Carolina at Chapel Hill, the Inter-University Centre for Astronomy and Astrophysics (IUCAA) at Pune, India, and the National Central University, Taiwan. The Murty family feels very happy to have added a small value to
this important project. Robo-AO KP is also supported by grants from the John Templeton Foundation and the Mt. Cuba Astronomical Foundation. The Robo-AO instrument was developed with support from the National Science Foundation under grants AST-0906060, AST-0960343, and AST-1207891, IUCAA, the Mt. Cuba Astronomical Foundation, and by a gift from Samuel Oschin. These data are based on observations at Kitt Peak National Observatory, National Optical Astronomy Observatory (NOAO Prop. ID: 15B-3001), which is operated by the Association of Universities for Research in Astronomy (AURA) under cooperative agreement with the National Science Foundation.

This research has made use of the Exoplanet Follow-up Observation Program website, which is operated by the California Institute of Technology, under contract with the National Aeronautics and Space Administration under the Exoplanet Exploration Program

This research has made use of the NASA Exoplanet Archive, which is operated by the California Institute of Technology, under contract with the National Aeronautics and Space Administration under the Exoplanet Exploration Program.

{\it Facilities:} \facility{PO:1.5m (Robo-AO)}, \facility{KPNO:2.1m	(Robo-AO)}

\bibliography{references}

\LongTables
\clearpage

\begin{deluxetable*}{lcccccccc}
\tablewidth{0pt}
\tablecaption{\label{tab:newkois_table}Detections of Objects Nearby 532 Previously Unpublished \textit{Kepler} Planet Candidates}
\tablehead{
\colhead{KOI} & \colhead{$\rm m_{\textit{Kep}}$} & \colhead{ObsID} & \colhead{Filter} & \colhead{Det. Significance} & \colhead{Separation} & \colhead{P.A.} & \colhead{Mag. Diff.} & \colhead{Previous} \\
 & \colhead{(mag)} &  & & \colhead{$\sigma$} & \colhead{($\arcsec$)} & \colhead{(deg.)} & \colhead{(mag)} &  \colhead{high res.?}  \\
}
\startdata
120 & 12.003 & 2016 Jul 15 & LP600 & 7.3 & 1.62$\pm$0.06 & 129$\pm$2 & 0.51$\pm$0.01 & \\ 
129 & 13.224 & 2016 Jul 15 & LP600 & 6.7 & 2.1$\pm$0.06 & 221$\pm$3 & 5.87$\pm$0.1 & \\ 
215 & 14.708 & 2016 Jul 14 & LP600 & 11.1 & 2.98$\pm$0.06 & 22$\pm$2 & 2.34$\pm$0.02 & \\ 
229 & 14.720 & 2016 Jun 17 & LP600 & 6.2 & 1.66$\pm$0.06 & 264$\pm$2 & 0.99$\pm$0.04 & \\ 
472 & 15.000 & 2016 Jun 18 & LP600 & 7.2 & 1.12$\pm$0.06 & 206$\pm$2 & 0.72$\pm$0.07 & \\ 
506 & 14.731 & 2016 Jun 18 & LP600 & 4.5 & 3.15$\pm$0.06 & 39$\pm$4 & 5.0$\pm$0.14 & \\ 
506 & 14.731 & 2016 Jun 18 & LP600 & 13.2 & 1.13$\pm$0.06 & 15$\pm$3 & 3.35$\pm$0.07 & \\ 
636 & 13.252 & 2016 Jun 17 & LP600 & 60.1 & 3.8$\pm$0.06 & 343$\pm$5 & 6.0$\pm$0.06 & \\ 
799 & 15.279 & 2016 Jun 16 & LP600 & 7.4 & 1.28$\pm$0.06 & 108$\pm$2 & 1.73$\pm$0.05 & \\ 
840 & 15.028 & 2016 Jun 17 & LP600 & 5.4 & 2.97$\pm$0.06 & 302$\pm$3 & 3.42$\pm$0.03 & \\ 
840 & 15.028 & 2016 Jun 17 & LP600 & 11.9 & 3.2$\pm$0.06 & 334$\pm$2 & 2.24$\pm$0.01 & \\ 
903 & 15.813 & 2016 Jun 27 & LP600 & 5.1 & 2.24$\pm$0.06 & 99$\pm$2 & 1.84$\pm$0.1 & \\ 
927 & 15.453 & 2016 Jul 08 & LP600 & 4.8 & 1.01$\pm$0.06 & 294$\pm$3 & 2.63$\pm$0.04 & \\ 
944 & 15.361 & 2016 Jul 08 & LP600 & 5.7 & 1.14$\pm$0.06 & 155$\pm$2 & 2.55$\pm$0.05 & \\ 
959 & 13.102 & 2016 Jun 25 & LP600 & 5.5 & 0.68$\pm$0.06 & 117$\pm$4 & 1.25$\pm$0.02 & F17\\ 
980 & 10.376 & 2016 Jul 15 & LP600 & 5.3 & 1.01$\pm$0.06 & 31$\pm$3 & 1.65$\pm$0.03 & F17\\ 
1075 & 13.056 & 2016 Jul 02 & LP600 & 4.3 & 1.07$\pm$0.06 & 93$\pm$3 & 2.63$\pm$0.07 & \\ 
1126 & 15.259 & 2016 Jun 19 & LP600 & 8.1 & 1.85$\pm$0.06 & 302$\pm$2 & 2.89$\pm$0.04 & \\ 
1188 & 15.381 & 2016 Jun 17 & LP600 & 11.1 & 3.39$\pm$0.06 & 202$\pm$2 & 2.16$\pm$0.02 & \\ 
1191 & 15.240 & 2016 Jun 17 & LP600 & 8.6 & 3.91$\pm$0.06 & 226$\pm$3 & 3.89$\pm$0.14 & \\ 
1254 & 12.777 & 2016 Jun 25 & LP600 & 5.0 & 2.98$\pm$0.06 & 28$\pm$2 & 0.88$\pm$0.05 & \\ 
1287 & 15.910 & 2016 Jun 18 & LP600 & 8.4 & 2.64$\pm$0.06 & 339$\pm$2 & 1.47$\pm$0.01 & \\ 
1450 & 13.480 & 2016 Jun 25 & LP600 & 5.4 & 1.74$\pm$0.06 & 208$\pm$2 & 2.46$\pm$0.04 & \\ 
1545 & 15.169 & 2016 Jun 18 & LP600 & 4.3 & 2.51$\pm$0.06 & 180$\pm$4 & 5.06$\pm$0.18 & \\ 
1614 & 11.413 & 2016 Jun 18 & LP600 & 8.7 & 3.37$\pm$0.06 & 87$\pm$2 & -0.44$\pm$0.06 & F17\\ 
1661 & 11.510 & 2016 Jun 18 & LP600 & 7.0 & 1.37$\pm$0.06 & 46$\pm$2 & 3.22$\pm$0.08 & \\ 
1729 & 15.424 & 2016 Jun 22 & LP600 & 10.2 & 3.83$\pm$0.06 & 210$\pm$3 & 3.81$\pm$0.07 & \\ 
1855 & 14.782 & 2016 Jun 17 & LP600 & 4.7 & 1.5$\pm$0.06 & 222$\pm$4 & 5.79$\pm$0.73 & \\ 
1901 & 13.340 & 2016 Jun 15 & LP600 & 20.2 & 3.82$\pm$0.06 & 105$\pm$2 & 2.16$\pm$0.02 & \\ 
1943 & 13.377 & 2016 Jun 22 & LP600 & 7.2 & 1.42$\pm$0.06 & 302$\pm$3 & 1.42$\pm$0.03 & \\ 
2880 & 15.918 & 2016 Jul 02 & LP600 & 6.8 & 3.39$\pm$0.06 & 257$\pm$ & 1.15$\pm$0.12 & \\ 
2897 & 15.361 & 2016 Jun 18 & LP600 & 8.4 & 2.65$\pm$0.06 & 200$\pm$2 & 2.98$\pm$0.04 & \\ 
3002 & 13.256 & 2016 Jun 17 & LP600 & 5.8 & 0.84$\pm$0.06 & 267$\pm$4 & 2.02$\pm$0.05 & \\ 
3156 & 7.899 & 2016 Jun 17 & LP600 & 9.5 & 1.24$\pm$0.06 & 203$\pm$2 & 2.09$\pm$0.03 & F17 \\ 
3156 & 7.899 & 2016 Jun 17 & LP600 & 15.8 & 3.06$\pm$0.06 & 288$\pm$3 & 5.02$\pm$0.06 & \\ 
3435 & 15.259 & 2016 Jun 16 & LP600 & 6.7 & 3.06$\pm$0.06 & 160$\pm$2 & 1.33$\pm$0.02 & \\ 
3435 & 15.259 & 2016 Jun 16 & LP600 & 14.9 & 3.52$\pm$0.06 & 301$\pm$2 & 0.58$\pm$0.04 & \\ 
3790 & 18.590 & 2016 Jun 22 & LP600 & 6.5 & 1.28$\pm$0.06 & 138$\pm$3 & 1.91$\pm$0.05 & \\ 
3805 & 11.356 & 2016 Jun 17 & LP600 & 3.8 & 3.7$\pm$0.06 & 199$\pm$4 & 5.37$\pm$0.02 & L12\\ 
3813 & 14.113 & 2016 Jun 18 & LP600 & 10.3 & 2.54$\pm$0.06 & 283$\pm$3 & 4.58$\pm$0.07 & \\ 
3813 & 14.113 & 2016 Jun 18 & LP600 & 7.8 & 2.13$\pm$0.06 & 58$\pm$4 & 4.22$\pm$0.02 & \\ 
3856 & 13.493 & 2016 Jun 16 & LP600 & 8.2 & 2.54$\pm$0.06 & 101$\pm$2 & 3.27$\pm$0.02 & \\ 
4368 & 13.046 & 2016 Jun 18 & LP600 & 12.3 & 2.33$\pm$0.06 & 162$\pm$3 & 3.28$\pm$0.01 & \\ 
4625 & 15.877 & 2016 Jul 15 & LP600 & 7.0 & 1.22$\pm$0.06 & 69$\pm$3 & 0.28$\pm$0.02 & \\ 
4707 & 11.660 & 2016 Jun 15 & LP600 & 11.4 & 3.7$\pm$0.06 & 13$\pm$2 & 6.41$\pm$0.03 & \\ 
4764 & 15.809 & 2016 Jun 19 & LP600 & 24.2 & 3.83$\pm$0.06 & 204$\pm$2 & 2.42$\pm$0.04 & \\ 
4793 & 15.374 & 2016 Jun 19 & LP600 & 5.6 & 2.37$\pm$0.06 & 225$\pm$4 & 4.25$\pm$0.09 & \\ 
5274 & 12.746 & 2016 Jun 22 & LP600 & 5.4 & 3.95$\pm$0.06 & 272$\pm$3 & 4.13$\pm$0.01 & \\ 
5426 & 13.701 & 2016 Jun 25 & LP600 & 4.8 & 2.93$\pm$0.06 & 152$\pm$2 & 1.75$\pm$0.11 & \\ 
5454 & 14.150 & 2016 Jun 25 & LP600 & 4.5 & 2.07$\pm$0.06 & 286$\pm$2 & 1.77$\pm$0.05 & \\ 
5475 & 13.093 & 2016 Jun 27 & LP600 & 9.1 & 3.19$\pm$0.06 & 70$\pm$3 & 3.65$\pm$0.01 & F17\\ 
5527 & 14.174 & 2016 Jul 02 & LP600 & 9.3 & 2.85$\pm$0.06 & 236$\pm$2 & 2.63$\pm$0.01 & \\ 
5552 & 13.344 & 2016 Jun 27 & LP600 & 4.8 & 1.09$\pm$0.06 & 165$\pm$3 & 0.82$\pm$0.04 & \\ 
5640 & 12.038 & 2016 Jun 28 & LP600 & 5.2 & 0.53$\pm$0.06 & 113$\pm$4 & 2.26$\pm$0.05 & \\ 
5672 & 14.333 & 2016 Jul 08 & LP600 & 10.1 & 3.17$\pm$0.06 & 169$\pm$3 & 4.58$\pm$0.19 & \\ 
5790 & 15.518 & 2016 Jun 28 & LP600 & 8.9 & 3.69$\pm$0.06 & 357$\pm$2 & -0.67$\pm$0.01 & \\ 
5792 & 15.705 & 2016 Jul 13 & LP600 & 11.1 & 3.59$\pm$0.06 & 116$\pm$2 & -0.07$\pm$0.07 & \\ 
5797 & 12.220 & 2016 Jul 13 & LP600 & 13.6 & 3.62$\pm$0.06 & 103$\pm$2 & 1.37$\pm$0.05 & \\ 
5868 & 13.787 & 2016 Jul 02 & LP600 & 3.5 & 2.8$\pm$0.06 & 94$\pm$2 & 2.71$\pm$0.09 & \\ 
5895 & 15.337 & 2016 Jul 15 & LP600 & 6.5 & 2.34$\pm$0.06 & 249$\pm$2 & 3.41$\pm$0.03 & \\ 
5941 & 13.783 & 2016 Jul 13 & LP600 & 6.7 & 1.07$\pm$0.06 & 216$\pm$4 & 5.28$\pm$0.22 & \\ 
5961 & 15.053 & 2016 Jul 13 & LP600 & 6.9 & 0.87$\pm$0.06 & 112$\pm$3 & 1.45$\pm$0.03 & \\ 
5993 & 12.873 & 2016 Jun 14 & LP600 & 7.8 & 1.25$\pm$0.06 & 217$\pm$3 & 3.06$\pm$0.04 & \\ 
6104 & 14.708 & 2016 Jun 16 & LP600 & 8.7 & 1.84$\pm$0.06 & 206$\pm$3 & 4.01$\pm$0.01 & \\ 
6224 & 12.962 & 2016 Jul 13 & LP600 & 9.0 & 2.97$\pm$0.06 & 167$\pm$2 & 4.19$\pm$0.01 & \\ 
6256 & 15.729 & 2016 Jun 15 & LP600 & 6.8 & 3.05$\pm$0.06 & 103$\pm$2 & 2.27$\pm$0.05 & \\ 
6297 & 14.043 & 2016 Jun 16 & LP600 & 10.1 & 2.56$\pm$0.06 & 103$\pm$2 & 1.55$\pm$0.01 & \\ 
6297 & 14.043 & 2016 Jun 16 & LP600 & 6.3 & 2.96$\pm$0.06 & 308$\pm$5 & 5.89$\pm$0.24 & \\ 
6384 & 15.992 & 2016 Jun 16 & LP600 & 13.1 & 3.53$\pm$0.06 & 285$\pm$2 & 2.09$\pm$0.01 & \\ 
6390 & 13.961 & 2016 Jun 17 & LP600 & 10.9 & 2.82$\pm$0.06 & 309$\pm$2 & 1.57$\pm$0.03 & \\ 
6600 & 10.715 & 2016 Jun 17 & LP600 & 15.2 & 2.36$\pm$0.06 & 315$\pm$4 & 5.28$\pm$0.02 & F17\\ 
6697 & 13.678 & 2016 Jun 22 & LP600 & 30.7 & 3.91$\pm$0.06 & 313$\pm$3 & 3.4$\pm$0.01 & \\ 
6783 & 15.472 & 2016 Jun 22 & LP600 & 5.5 & 3.25$\pm$0.06 & 178$\pm$3 & 3.31$\pm$0.06 & \\ 
6793 & 14.835 & 2016 Jun 22 & LP600 & 4.8 & 2.84$\pm$0.06 & 309$\pm$4 & 4.47$\pm$0.25 & \\ 
6835 & 13.903 & 2016 Jun 22 & LP600 & 8.2 & 3.08$\pm$0.06 & 78$\pm$5 & 5.34$\pm$0.07 & \\ 
6907 & 15.930 & 2016 Jun 22 & LP600 & 15.1 & 3.35$\pm$0.06 & 99$\pm$2 & -0.36$\pm$0.07 & \\ 
6918 & 14.596 & 2016 Jun 22 & LP600 & 5.2 & 0.62$\pm$0.06 & 98$\pm$4 & 1.33$\pm$0.04 & \\ 
7002 & 14.991 & 2016 Jun 27 & LP600 & 6.2 & 3.2$\pm$0.06 & 247$\pm$2 & 2.95$\pm$0.06 & \\ 
7003 & 14.106 & 2016 Jun 25 & LP600 & 4.8 & 3.78$\pm$0.06 & 285$\pm$2 & 1.9$\pm$0.09 & \\ 
7032 & 12.646 & 2016 Jun 22 & LP600 & 4.1 & 2.74$\pm$0.06 & 182$\pm$4 & 5.8$\pm$0.13 & F17\\ 
7050 & 13.506 & 2016 Jun 27 & LP600 & 5.3 & 1.78$\pm$0.06 & 129$\pm$2 & 2.5$\pm$0.01 & \\ 
7087 & 12.457 & 2016 Jul 02 & LP600 & 5.1 & 1.89$\pm$0.06 & 165$\pm$2 & 1.69$\pm$0.02 & \\ 
7129 & 13.839 & 2016 Jun 28 & LP600 & 6.2 & 1.27$\pm$0.06 & 191$\pm$2 & 2.39$\pm$0.02 & \\ 
7220 & 15.102 & 2016 Jul 12 & LP600 & 9.9 & 3.57$\pm$0.06 & 27$\pm$2 & 1.33$\pm$0.07 & \\ 
7389 & 12.148 & 2016 Jul 14 & LP600 & 9.8 & 1.84$\pm$0.06 & 291$\pm$6 & 6.2$\pm$0.04 & \\ 
7408 & 15.949 & 2016 Jul 14 & LP600 & 5.8 & 1.67$\pm$0.06 & 184$\pm$2 & 2.65$\pm$0.06 & \\ 
7455 & 11.419 & 2016 Jul 08 & LP600 & 6.1 & 1.86$\pm$0.06 & 306$\pm$3 & 2.39$\pm$0.03 & F17\\ 
7470 & 13.870 & 2016 Jul 15 & LP600 & 6.4 & 1.52$\pm$0.06 & 303$\pm$2 & 0.17$\pm$0.02 & F17\\ 
7501 & 11.308 & 2016 Jul 15 & LP600 & 5.6 & 1.15$\pm$0.06 & 15$\pm$2 & 1.36$\pm$0.06 & \\ 
7527 & 13.573 & 2016 Jul 15 & LP600 & 6.4 & 2.75$\pm$0.06 & 98$\pm$4 & 4.41$\pm$0.06 & \\ 
7539 & 14.813 & 2016 Jul 15 & LP600 & 7.3 & 2.97$\pm$0.06 & 348$\pm$3 & 3.03$\pm$0.03 & \\ 
7540 & 13.852 & 2016 Jul 14 & LP600 & 8.5 & 3.67$\pm$0.06 & 152$\pm$4 & 5.64$\pm$0.06 & \\ 
7546 & 12.667 & 2016 Jun 17 & LP600 & 6.7 & 2.93$\pm$0.06 & 223$\pm$5 & 5.92$\pm$0.07 & \\ 
7572 & 9.748 & 2016 Jun 22 & LP600 & 7.5 & 2.97$\pm$0.06 & 16$\pm$4 & 5.14$\pm$0.01 & 
\enddata
\tablecomments{
References for previous high-resolution observations are denoted using the following codes:   \citealt{lillo12} (L12), \citealt{furlan16} (F17)
}
\end{deluxetable*}

\clearpage

\begin{appendix}
\section{PSF Subtraction Collisions}
\label{sec:psfcollisions}

By using other Robo-AO observations of KOIs as reference images, there is a possibility that an image used as a reference PSF will have a nearby star at a similar position with respect to its host star as the image being modeled. Only companions at separations less than 1\arcsec could potentially avoid detection by both our visual search and the automated companion detection routine. Such a scenario (a ``collision'') could lead to real companions being removed from target images if they coincide with a reference star's companion. To estimate how near to each other the companions must be for a collision to occur, we ran the PSF subtraction routine on a set of ten targets which have detected nearby stars at varying separations within 1\arcsec. We then include a copy of each target image as one of the reference PSFs. In each case, the nearby star is not detected in the subtracted image by eye or by the automated companion detection routine with a significance $>$3$\sigma$. The reference image is then rotated by two degrees, and the PSF subtraction routine is rerun. This process is iterated until the nearby star is able to be detected in the subtracted image. We find on average the companion in the reference image must be within 0\farcs05 of the position of the nearby star in the original image for a collision to occur.

To estimate the expected number of collisions in our analysis, we use the observed distribution of nearby stars from our survey to populate a simulated KOI survey. For each nearby star detected with separations less than 1\arcsec, we randomly drew twenty other reference stars. We counted every time a reference star fell within 0\farcs05 of the original star as a collision. With 100 simulations performed, we estimate the number of expected companions missed in our survey due to collisions is 0.44$\pm$0.18, or approximately one every two surveys.

The visual search for companions, however, will greatly reduce the number of expected companions missed in our analysis. Within our observations, we find two potential collisions (KOIs 3497 and 4098, and KOIs 6202 and 6602). Neither of these sets of colliding images were used as a reference image for each other in the initial data analysis. We reran the PSF subtraction routine for both sets using the colliding system for each as a reference image. In each case, the nearby star is only partially subtracted and is still detectable within the subtracted image. This suggests that slight alterations in the Robo-AO PSFs are sufficient to effectively eliminate the possibility that a real companion will be erroneously subtracted off by the PSF subtraction routine.

\section{Updated Planetary Radii}
In Table$~\ref{tab:radii}$, we derive the corrected planetary radii for every \textit{Kepler} planetary candidate with a detected nearby star (as described in Section \ref{sec:implications}).

\LongTables
\tabletypesize{\scriptsize}

\begin{deluxetable*}{ccccccccc}
\tablewidth{0pt}
\tablecaption{\label{tab:radii}Implications on Derived Radius of \textit{Kepler} Planetary Candidates}
\tablehead{
\colhead{Object} & \colhead{Sep.} & \colhead{$\Delta$m} & \colhead{Reference\textsuperscript{a}} & \colhead{R$_{\star, target}$\textsuperscript{b}} & \colhead{R$_{\star, secondary}$\textsuperscript{c}} & \colhead{R$_{p,0}$\textsuperscript{d}} & \colhead{R$_{p,prim.}$\textsuperscript{e}} & \colhead{R$_{p,sec.}$\textsuperscript{f}}\\
\colhead{} & \colhead{(\arcsec)} & \colhead{(mags)} & & \colhead{(R$_{\odot}$)} & \colhead{(R$_{\odot}$)} & \colhead{(R$_{\Earth}$)} & \colhead{(R$_{\Earth}$)} & \colhead{(R$_{\Earth}$)}
}
\startdata
0001.01 & 1.13 & 3.95 & P1 & 0.96 & 0.58 & 13.0 & 13.2 & 49.1  \\ 
0004.01 & 3.42 & 4.46 & P2 & 2.99 & 0.58 & 12.9 & 13.0 & 19.8  \\ 
0013.01 & 1.16 & 0.19 & P1 & 3.03 & 0.58 & 21.4 & 29.1 & 6.1  \\ 
0042.01 & 1.74 & 3.04 & P2 & 1.34 & 0.75 & 2.43 & 2.5 & 5.7  \\ 
0044.01 & 3.42 & 4.03 & TW & 1.07 & 0.6 & 12.0 & 12.1 & 43.5  \\ 
0070.01 & 3.86 & 5.74 & TW & 0.93 & 0.56 & 2.99 & 3.0 & 25.5  \\ 
0070.02 & 3.86 & 5.74 & TW & 0.93 & 0.56 & 1.78 & 1.78 & 15.2  \\ 
0070.03 & 3.86 & 5.74 & TW & 0.93 & 0.56 & 2.59 & 2.6 & 22.1  \\
\enddata
\tablecomments{
\tablenotetext{a}{Reference for nearby star detection: \citep[P1,][]{law14}, \citep[P2,][]{baranec16}, \citep[P3,][]{ziegler16}, (TW, this work).}
\tablenotetext{b}{Primary stellar radius estimate from \citet{dr25}.}
\tablenotetext{c}{Estimated radius of secondary stellar companion in the scenario where it is bound to the primary star, using absolute magnitude difference in the \textit{Kepler} band and the Dartmouth stellar models \citep{dotter08}.}
\tablenotetext{d}{Original planetary radius estimate, from NASA Exoplanet Archive.}
\tablenotetext{e}{Estimated eclipsing object radius in the scenario where it is physically bound to the target star, corrected for transit dilution caused by the presence of nearby stars.}
\tablenotetext{f}{Estimated eclipsing object radius in the scenario where it is bound to the companion star, correcting for transit dilution by nearby stars and using the stellar radius estimate of the companion in this table.}
}
\end{deluxetable*}

\section{Full Robo-AO Observations Table}

In Table$~\ref{tab:whitelist}$, we list KOIs observed with Robo-AO at Kitt Peak, including the date the target was observed, observation quality (as described in Section \ref{sec:imageperf}), and the presence of detected companions.

{\footnotesize
\tabcolsep=0.12cm
\begin{longtable}{ccccc}
\caption{Full Robo-AO Observation List \label{tab:whitelist}}\\
\hline
\hline
\noalign{\vskip 3pt}  
\text{KOI} & m\textit{$_{Kep}$} & \text{ObsID} & \text{Obs. Qual.} & \text{Comp.}\\ [0.1ex]
 & (mags) & &  &  \text{Det.?} \\ [0.1ex]
\hline
\noalign{\vskip 3pt}  
\endfirsthead
\multicolumn{5}{c}
{\tablename\ \thetable\ -- \textit{Continued}} \\
\hline \hline
\noalign{\vskip 3pt} 
\text{KOI} & m\textit{$_{Kep}$} & \text{ObsID} & \text{Obs. Qual.} & \text{Comp.}\\ [0.1ex]
 & (mags) & &  & \text{Det.?} \\ [0.1ex]
\hline
\noalign{\vskip 3pt}  
\endhead
\endfoot
\hline
\endlastfoot
K019 & 11.37 & 2016 Jun 14 & medium &  \\ 
K025 & 13.5 & 2016 Jul 08 & medium &  \\ 
K048 & 13.72 & 2016 Jun 25 & low &  \\ 
K072 & 10.96 & 2016 Jul 02 & high &  \\ 
K0120 & 12.0 & 2016 Jul 15 & medium & yes \\ 
K0125 & 13.78 & 2016 Jul 12 & medium &  \\ 
K0129 & 13.22 & 2016 Jul 15 & medium & yes \\ 
K0130 & 13.32 & 2016 Jun 18 & medium &  \\
\end{longtable}}

\end{appendix}
\end{document}